\documentclass[aps,superscriptaddress,showkeys,longbibliography,notitlepage]{revtex4-1}
\usepackage[usenames,dvipsnames]{xcolor}
\usepackage{latexsym}
\usepackage{amsmath}
\usepackage{amsfonts}
\usepackage{amssymb}
\usepackage{bbm}
\usepackage{epsfig,graphics,color,calc,graphicx,float}
\usepackage{color}
\usepackage{graphics}
\usepackage{graphicx}
\usepackage{times}

\newcommand{\be}{\begin{equation}}
\newcommand{\ee}{\end{equation}}

\newcommand{\bea}{\begin{eqnarray}}
\newcommand{\eea}{\end{eqnarray}}
\newcommand{\beas}[1]{\begin{subequations}\label{#1}\bea}
\newcommand{\eeas}{\eea\end{subequations}}

\newcommand{\hidemc}[1]{}

\newcommand{\mcrit}{m_\textrm{crit}}
\newcommand{\meq}{m_\textrm{eq}}
\newcommand{\betah}{\beta h}
\renewcommand{\eqref}[1]{Eq.\ {\color{OliveGreen}(\ref{#1})}}
\newcommand{\figref}[1]{{\color{MidnightBlue}Fig.\ \ref{#1}}}
\newcommand{\figsref}[2]{{\color{MidnightBlue}Figs.\ \ref{#1} and \ref{#2}}}
\newcommand{\figureref}[1]{{\color{MidnightBlue}Figure\ \ref{#1}}}
\newcommand{\skipbetah}{h}

\begin{document}

\title{Emergence of stationary uphill currents in 2D Ising models: the role of reservoirs and boundary conditions}

\author{Matteo Colangeli}
\email{matteo.colangeli1@univaq.it}
\affiliation{Dipartimento di Ingegneria e Scienze dell'Informazione e Matematica, Universit\`a degli Studi dell'Aquila, 67100 L'Aquila, Italy}
\author{Claudio Giberti}
\email{claudio.giberti@unimore.it}
\affiliation{Dipartimento di Scienze e Metodi dell'Ingegneria,
Universit\`a di Modena e Reggio E., \\ Via Amendola 2, Padiglione Morselli, I-42122 Reggio E., Italy.}
\author{Cecilia Vernia}
\email{cecilia.vernia@unimore.it}
\affiliation{Dipartimento di Scienze Fisiche Informatiche e Matematiche, Universit\`a  di Modena 
e Reggio E.,\\
Via Campi 213/B, I-41125 Modena, Italy\\}
\author{Martin Kr\"{o}ger}
\email{mk@mat.ethz.ch}
\affiliation{Polymer Physics, Department of Materials, ETH Zurich, CH--8093 Zurich, Switzerland}

\begin{abstract}
We investigate the dynamics of a 2D Ising model on a square lattice with conservative Kawasaki dynamics in the bulk, coupled with two external reservoirs that pull the dynamics out of equilibrium.  Two different mechanisms for the action of the reservoirs  are considered. In the first, called ISF, the condition of local equilibrium  between reservoir and the lattice is not satisfied. The  second mechanism, called DB, implements a detailed balance condition, thus satisfying the local equilibrium property.  We provide numerical evidence that, for a suitable choice of the temperature (i.e. below the critical temperature of the equilibrium 2D Ising model)  and the reservoir magnetizations, in the long time limit the ISF model undergoes a ferromagnetic phase transition and gives rise to stationary \textit{uphill} currents, namely positive spins diffuse from the reservoir with lower magnetization to the reservoir with higher magnetization. The same phenomenon does not occur for DB dynamics with properly chosen boundary conditions. Our analysis extends the results reported in Colangeli \textit{et al.} ({\em Phys. Rev. E} \textbf{97}:030103(R), 2018), shedding also light on the effect of  temperature and the role of different boundary conditions for this model. These issues may be relevant in a variety of situations (e.g. mesoscopic systems) in which the violation of the local equilibrium condition produces unexpected phenomena that seem to contradict the standard laws of transport.
\end{abstract}

%\pacs{81.05.Uw,68.37.-d,73.20-r}

\maketitle

\tableofcontents{}

\section{\label{intro} Introduction}

The presence of uphill diffusion in particle or spin models is a remarkable problem of statistical mechanics, that was first envisaged by Darken in his pioneering work dating back to 1949 \cite{Darken49}. While Fick's law of diffusion dictates that particles diffuse \textit{against} the concentration gradient (\textit{downhill} diffusion), in presence of a multi-component system or an external field particles may be found migrating up the gradient \cite{CDMP17bis,CC17}, giving thus rise to the so-called \textit{uphill} currents, confirmed both experimentally and theoretically in numerous studies \cite{Lesher1994,Frink2000,Ferri2006,Kuhl2007,Yu2007,Sundman2009,Vielzeuf2011,Boudin2012,Rougier2013,Laeuerer2015}. Remarkably, in \cite{CDMP16,CDMP17} some numerical and theoretical evidence supported the conclusion that uphill diffusion may also occur in single-component systems in presence of a phase transition. In these works, the authors consider a 1D lattice gas model coupled to external particle reservoirs, in which particles that hop on the lattice are subject to an exclusion principle and are equipped with a long range, Kac-like interaction \cite{Presutti09}, which gives rise, at sufficiently low temperatures, to a phase transition. By adopting spin variables, the authors show that when the absolute value of the magnetization at the boundaries is larger than the equilibrium mean field magnetization, a sharp interface (called the ``instanton'' therein) appears at the center of the lattice accompanied by a downhill current. The novelty comes when the absolute magnetization at the boundary is lowered below the equilibrium mean field value: the interface, referred to as the ``bump'', is then observed moving towards one of the boundaries, and the current changes sign, namely positive spins move from the reservoir with higher magnetization towards the one with lower magnetization. One might consider, therefore, an experimental set-up in which two finite reservoirs, equipped with metastable values of magnetization, are connected by two channels: one crossed by an uphill current and another one (in which the long-range Kac interaction is absent) featuring a Fickian current. The result would indeed be a closed circuit in which, in presence of a phase transition, current spontaneously flows across the ring. This phenomenon was indeed observed and reproduced in \cite{CDMP17} by performing extensive Monte Carlo simulations of the model (note that no violation of the thermodynamic principles occurs therein, as the total energy of the system is not a conserved quantity). A step forward was then taken in \cite{CGGV18}, in which a similar phenomenology was also observed in a 2D Ising model on a square lattice coupled to two external magnetization reservoirs attached at the right and left boundaries (whose magnetizations are equal, respectively, to $m_+>0$ and $m_-=-m_+$). The presence of a stationary uphill current is induced by the reservoirs, whose updating mechanism at the boundaries breaks the condition of local detailed balance, cf. also \cite[Eq. 1.6]{ELS1990}. In particular, the authors show that, in presence of a ferromagnetic phase transition, a certain critical value $\mcrit$ marks the transition between two different regimes (referred to, in \cite{CGGV18}, as the \textit{stable} and \textit{metastable} one), and characterized, respectively by downhill and uphill currents, when $m_+$ is, respectively, larger or smaller than  $\mcrit$.

We recall that in the absence of reservoirs and in the infinite volume limit,
the equilibrium 2D Ising model undergoes a phase transition at the
inverse critical temperature \cite{Onsager1944}
\be
\label{betac}
\beta_c = \frac{\ln(1+\sqrt{2})}{2} \approx 0.440686\, .
\ee
For inverse temperatures $\beta>\beta_c$ the 2D Ising model (with vanishing external magnetic field) exhibits a spontaneous magnetization
given by \cite{Yang1952}
\be
\label{mbeta}
m_\beta = \left[1-\frac{1}{\sinh^{4}\left(2\beta\right)}\right]^{1/8}\, .
\ee

In \cite{CGGV18} it is claimed that  the critical value $\mcrit$, evaluated at $\beta=1$, can also be estimated by measuring the magnetization value $\meq$ (that approaches, in the large volume limit, the critical value $m_\beta$ in \eqref{mbeta} \cite{Onsager1944,Yang1952}) evaluated at the rightmost column of an Ising model in equilibrium conditions (i.e. in the absence of reservoirs and external magnetic fields) and characterized by a conservative dynamics.
The claim above thus indicates the possibility that a ``nonequilibrium'' observable, such as $\mcrit$, characterizing the dynamics of a boundary-driven Ising model, may be estimated from the analysis of an equilibrium Ising model. We shall tackle carefully this question in Section\ \ref{sec:results} and will show that the two quantities $\mcrit$ and $\meq$, evaluated for some fixed $L$ and $\beta$, are generally different from one another, their deviation becoming negligible only for large values of $\beta$ and by choosing suitable boundary conditions (b.c.s). At $\beta=1$ and with the b.c. considered in \cite{CGGV18}, the two observables are, indeed, almost coinciding.

In this work, we continue along the path traced in \cite{CGGV18} and aim to investigate in more detail the effect of different b.c.s for the considered 2D Ising model, cf. also \cite{Jan1983}, and unveil novel regimes other than the stable and metastable ones, by dropping further below the absolute value of the magnetization of the reservoirs.
In particular, we shall discuss the behavior of the \textit{equation of state}, namely the relation between the stationary current and the value of the magnetization at the boundaries, thus extending the preliminary results obtained in \cite{CGGV18}, cf. Fig. 2 therein.

The manuscript is organized as follows. The geometry and the bulk dynamics of our model are introduced in Section \ref{sec:model}, together with the boundary conditions in Section \ref{sec:bc}, and details of the spin-updating mechanisms (due to the coupling with the reservoirs) in Section \ref{sec:spin-update}. The role of the chosen spin-updating mechanism, and the comparison with a different mechanism based on detailed balance, are clarified in the same section. Definitions of observables and implementation details are provided in Sections \ref{sec:obs} and \ref{sec:implem}.  The results of our simulations are presented and discussed in Section \ref{sec:results}. Section \ref{sec:concl} is devoted to conclusions.

\section{The models}
\label{sec:model}

We consider non-equilibrium Monte Carlo (MC) dynamics of the nearest-neighbor ferromagnetic Ising model on a finite square 2D binary $L\times L$ lattice $\Lambda$ of linear size $L$, that is periodic only in one dimension (the vertical or synonymously, the $y$--direction). We shall denote by $\sigma_{i}\in\{-1,+1\}$ the ``bulk'' spin state at the lattice coordinate $i=(x,y)\in \Lambda$.

\begin{figure}[t]
\centering
\includegraphics[width=0.7\textwidth]{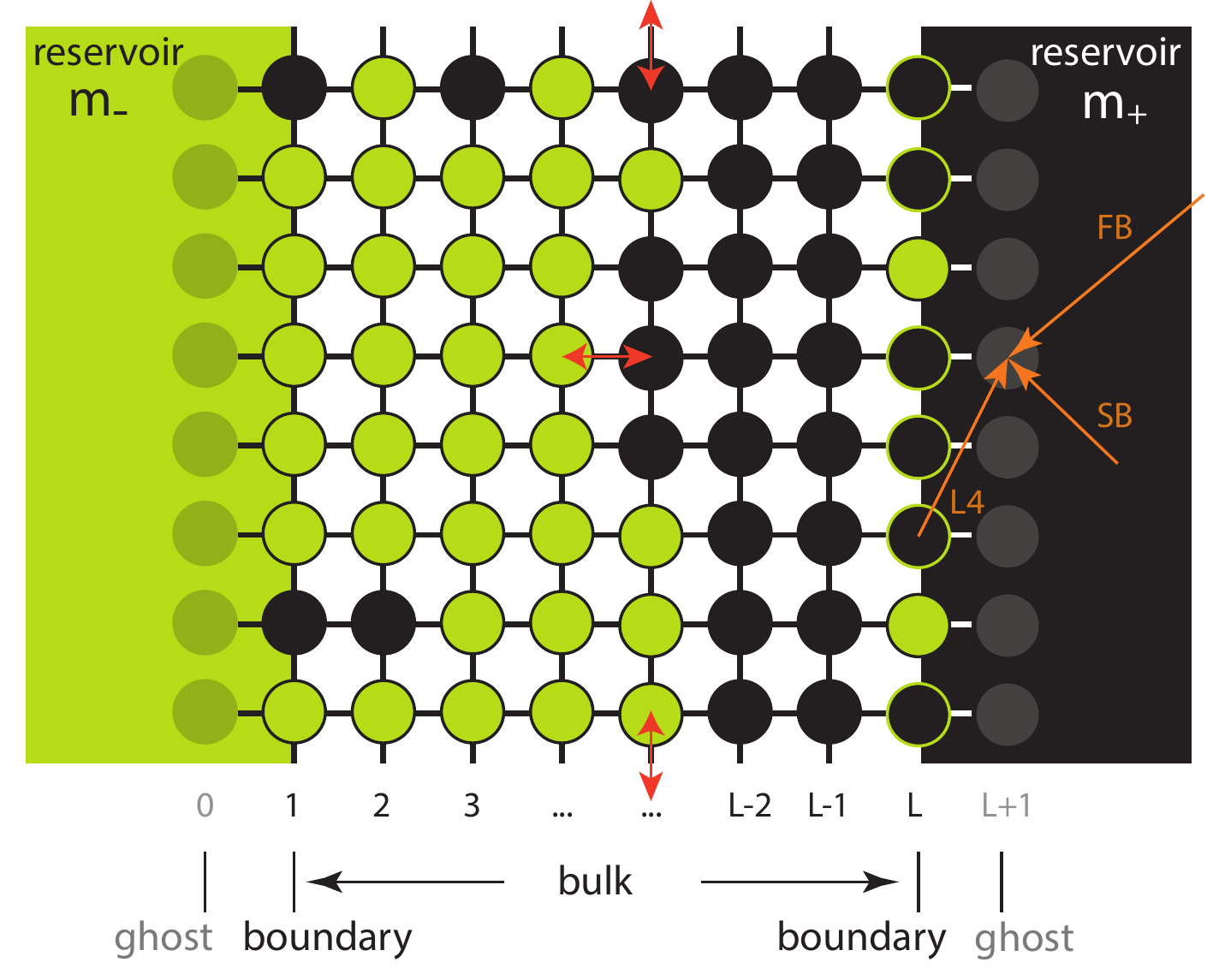}
\caption{Schematic picture of the 2D Ising model coupled to ideal reservoirs
at the left and right boundaries, characterized by the constant magnetizations, respectively, $m_-$ and $m_+$. Spins up and spins down are represented, respectively, with black and green circles, whereas the shaded circles in $x=0$ and $x=L+1$ represent the so-called \textit{ghost spins}, defining the b.c. of the model (they are not part, in general, of the reservoirs). The horizontal double-headed arrow at the center of the lattice means that a Kawasaki-type of dynamics holds in the bulk, whereas the vertical double-headed arrows at the bottom and at the top of the lattice recall that periodic b.c.s apply on the vertical direction. Finally, the single-headed arrows pointing towards one ghost spin in $x=L+1$ represent the different b.c.s considered for this model.}
\label{fig0}
\end{figure}

The system exhibits conservative exchange dynamics in the bulk (Section\ \ref{sec:bulk}), and is coupled to two infinite and interaction-free magnetization reservoirs on the horizontal direction (the $x$--direction, see \figref{fig0} for a description of the set-up), denoted as $\mathcal R_+$ and $\mathcal R_{-}$, located, respectively, at the right and the left ends of the lattice.
To this end, we define a bounded domain $\Lambda^o$, constituted by two vertical stripes, each made of $L$ lattice sites, located at $x=\bar{x}\in\{0,L+1\}$. We shall then denote by $\sigma^o_{i}$, $i=(\bar{x},y)\in \Lambda^o$  the ghost spins, picking up values in a set which depends on the chosen boundary model, as described in Section\ \ref{sec:bc}.
We denote by $N_b=2L^2+L$ the total number of bonds in the system,
consisting of $L^2+L(L-1)$ bulk bonds between adjacent spins in $\Lambda$ and $2L$ horizontal bonds connecting the spins at the boundaries of $\Lambda$ (hereafter called \textit{boundary spins}, located at $x=x_b\in\{1,L\}$) with the $2L$ so-called \textit{ghost spins} in $\Lambda^o$.
Note that the Ising model is equivalent to a lattice gas model via the standard
mapping between $L^2$ spin variables $\sigma_{i}$ and occupation variables $\eta_{i}=(1+ \sigma_{i})/2\in \{0,1\}$ with $\eta_{i} = 1$ (resp.  $\eta_{i} = 0$) denoting
the presence (resp. absence) of a particle.

Calling $\sigma=\{\sigma_i\}$, $i\in \Lambda$ and $\sigma^o=\{\sigma^o_i\}$, $i\in \Lambda^o$, respectively, the \textit{spin configuration} and  the \textit{boundary condition}, we define the Hamiltonian
\be
\label{H-Ising}
H(\sigma|\sigma^o) = - \frac{1}{2}\sum_{\underset{i,j\in \Lambda}{\langle i,j\rangle}} \sigma_i \sigma_j- \sum_{\underset{i\in \Lambda, j\in \Lambda^o}{\langle i,j\rangle}} \!\!\!\sigma_i \sigma^o_j,
\ee
where $\langle i,j\rangle$ denotes a nearest neighbor pair, and the magnetization
of spin configuration $\sigma$ is given by
\be
\label{magn}
 m(\sigma) = \frac{1}{L^2} \sum_{i\in\Lambda} \sigma_i.
\ee

Both the properties of the ghost spins at $\bar{x}$ (Section\ \ref{sec:bc}) and the dynamics of the boundary spins at $x_b$ (Section\ \ref{sec:spin-update}) are eventually affected by the two reservoirs $\mathcal R_+$ and $\mathcal R_{-}$. These reservoirs are fully characterized by their constant magnetizations $m_\textrm{res}$, respectively $m_+ \in [0,1]$ and  $m_-=-m_+$, respectively. In the sequel, when convenient, we will use
\be
 \betah_\textrm{res}=\tanh^{-1}(m_\textrm{res}) = \frac{1}{2}\ln \left(\frac{1+m_\textrm{res}}{1-m_\textrm{res}}\right), \qquad \textrm{m}_\textrm{res}\in\{m_-,m_+\}
 \label{defbetah}
\ee
in which $h_\textrm{res}$ can be regarded as the strength of a fictitious, dimensionless magnetic field, taking values $h_+\ge0$ and $h_-=-h_+$ at the left and the right reservoirs, respectively.
It will become soon clear, in Section\ \ref{sec:spin-update}, the reason of this notation: when the mechanism of interaction with the reservoirs is based on \textit{detailed balance}, $\tanh^{-1}(m_+)$ attains indeed the structure of a dimensionless magnetic field multiplied by inverse temperature $\beta$.
Table\ \ref{Table1} provides the translation from $m_+$ values to $\betah_+$ values considered in most plots of the following sections.

\begin{table}[tb]
\centering
\begin{tabular}{r@{\quad}r@{\qquad\qquad}r@{\quad}r@{\qquad\qquad}r@{\quad}r@{\qquad\qquad}r@{\quad}r}
\hline\hline
$m_+$ & $\betah_+$ & $m_+$ & $\betah_+$ & $m_+$ & $\betah_+$ & $m_+$ & $\betah_+$\\
\hline
 0.000000 & 0.00 &
 0.716298 & 0.90 &
 0.874053 & 1.35 &
0.998778 & 3.70 \\
 0.148885 & 0.15 &
 0.781806 & 1.05 &
0.970452 & 2.10 &
0.998894 & 3.75 \\
 0.291313 & 0.30 &
 0.817754 & 1.15 &
0.978026 & 2.25 &
0.999329 & 4.00 \\
 0.537050 & 0.60 &
 0.833655 & 1.20 &
0.998650 & 3.65 &
0.999955 & 5.35 \\
 0.604368 & 0.70 &
 0.848284 & 1.25 &
0.999273 & 3.96 &
1.000000 & $\infty$ \\
\hline\hline
\end{tabular}
\caption{
The sorted list of $(m_+,\betah_+)$ pairs contains all values 'selected' in this manuscript.}
\label{Table1}
\end{table}

\subsection{Kawasaki dynamics in the bulk}
\label{sec:bulk}

We consider the Ising system with reservoirs at inverse temperature $\beta$ and let the spins evolve
following a continuous-time stochastic dynamics with two contributions: a conservative exchange
dynamics in the bulk and and spin flip mechanisms at the opposing vertical boundaries.
%\textbf{independent spin flip} (ISF) mechanism at the opposing vertical boundaries.\\
More precisely, in the  bulk, the spins follow a Kawasaki dynamics,
i.e. the two spins of a selected bond $\langle i,j\rangle$, with $i,j\in \Lambda$, exchange (``bond flip'') their values with rate
\bea
\label{kawa}
 c_{i,j}^\textrm{bulk} = \textrm{min}(1,e^{-\beta \Delta H}), \qquad \Delta H = H(\sigma^{ij}|\sigma^o) - H(\sigma|\sigma^o),
\eea
where
$\sigma^{ij}$ denotes the configuration obtained from $\sigma$ by
exchanging the spins at sites $i$ and $j$. In the sequel we shall investigate the stationary dynamics corresponding to two different spin-updating mechanisms at the two vertical boundaries due the interaction with the reservoirs $\mathcal R_-$ and $\mathcal R_{+}$, and three different b.c.s.

\subsection{Boundary conditions}
\label{sec:bc}

Periodic b.c.s hold along the vertical direction of the considered model. The focus, hereafter, will be the investigation of different b.c.s imposed along the horizontal direction. These will indeed be found to affect the stationary magnetization profiles $\overline{m}_x$ (as function of grid coordinate $x$)
as well as the stationary, $x$-independent current $J$,
which is defined as the time-averaged rate of change of a boundary spin value.

The ``L4'' b.c., inspired by \cite{BP2003} and recently implemented in \cite{CGGV18}, is defined as follows
\begin{equation}
\mathrm{{\bf [L4]}}\qquad \sigma_{(\bar{x},y)}^o = \sigma_{(x_b,y-L/4)},
%\quad \textrm{with}\quad |\bar{x}-x_b| = 1 \qquad
\qquad \forall {\bar{x}\in\{x_b\pm 1\}}, \forall {y\in\{1,..,L\}}.
\end{equation}
The distance $L/4$ was chosen so as to make the two spins $ \sigma_{(x_b,y)}$ and $ \sigma_{(x_b,y-L/4)}$ sufficiently uncorrelated.
The second b.c. we consider is the so-called ``free boundary'' (FB) condition, in which one takes
\begin{equation}
\mathrm{{\bf [FB]}}\qquad\sigma_{i}^o=0, \quad \forall {i\in\Lambda^o}\,.
\end{equation}
Note that in both the L4 and FB b.c.s, the properties of the reservoirs do not enter directly.
The third b.c. we shall consider will be referred to, hereafter, as the ``stochastic boundary'' (SB) condition. The SB rule dictates that the probability
that a ghost spin takes a certain value is ruled by the magnetization $m_\textrm{res}$ of the reservoir next to $\sigma_i^o\in\{-1,+1\}$, by assuming that the average value of $\sigma_i^o$ is $m_\textrm{res}$, and reads

\begin{equation}
\label{ghsp}
\mathrm{{\bf [SB]}}\qquad\textrm{prob}(\sigma_{i}^o) =
\frac{1 + \sigma_{i}^o m_\textrm{res}}{2}, \quad \forall i\in\Lambda^o\, .
\end{equation}
where $m_\textrm{res} \in \{m_-,m_+\}$ is the magnetization of the reservoir close to $x=x_b$.

\subsection{Spin-updating mechanisms}
\label{sec:spin-update}

We shall now consider two different spin-updating mechanism at the vertical boundaries of the lattice, that mimic the action of the two reservoirs on the system. As shown in \figref{fig0}, the system is coupled, to its horizontal boundaries, to the two  reservoirs $\mathcal R_+$ and $\mathcal R_{-}$, having magnetization equal, respectively, to $m_+ \in [0,1]$ and  $m_-=-m_+$.

The first mechanism we shall consider in this work is the one considered in \cite{CGGV18}, called \textit{independent spin flip} (ISF).
According to the ISF mechanism, a selected boundary spin $\sigma_{i}$, $i=(x_b,y)$ is updated to its new value $\sigma'_{i}\in\{-1,1\}$ (which may coincide with $\sigma_{i}$) with a rate $c^\textrm{ISF}$ which is independent from the current state $\sigma_i$ and only dictated by the magnetization of the adjacent reservoir, i.e.:
\begin{equation}
\label{isf}
%c_\textrm{res}(\sigma'_{i}=\pm 1) = \frac{1\pm m_\textrm{res}}{2}
%c_i^\textrm{ISF}(\sigma'_{i}=\pm 1) = \frac{1\pm m_\textrm{res}}{2}
c^\textrm{ISF}(\sigma'_{i}) = \frac{1 + \sigma_i' m_\textrm{res}}{2}
\end{equation}
where $m_\textrm{res} \in \{m_-,m_+\}$ is the magnetization of the reservoir close to $x=x_b$.
We thus flip $\sigma_i$ with probability $(1-\sigma_i m_\textrm{res})/2$.
Note further that each reservoir plays a double role with the ISF-SB dynamics:
(i) it updates the boundary spins $\sigma_i\rightarrow \sigma'_i$, $i=(x_b,y)\in\Lambda$ with a rate defined by \eqref{isf},
(ii) it acts as a real ``boundary'' surrounding the lattice $\Lambda$ by fixing the stochastic b.c. $\sigma^o$ (i.e. the stochastic character of all the ghost spins in $\Lambda^o$), according to \eqref{ghsp}.

A second, so-called NSF spin-updating mechanism is obtained by considering a model that is fully decoupled from the reservoirs, namely no spin flip (NSF) mechanism takes place at the leftmost and rightmost vertical boundaries of the lattice.
For this model, the bulk dynamics is still defined by \eqref{kawa}, but \eqref{isf} is replaced by
\begin{equation}
\label{nsf}
c^\textrm{NSF}=0\, .
\end{equation}
Since reservoirs are not part of the description for the NSF dynamics, the SB b.c.
as defined in Section\ \ref{sec:bc} is no longer relevant here. However, we can still define the SB b.c. in the NSF case by considering \eqref{ghsp}  with a given $m_+$ (and $m_-=-m_+$)
even if these values are not associated to a reservoir.
It is readily seen that the NSF dynamics conserves, at any time step, the total magnetization, that is fixed by the initial configuration.
The presence of a spin-updating mechanism induced by the reservoirs, or the lack thereof, affects significantly the critical value of magnetization corresponding to a regime of vanishing current, as it will be seen in Section\ \ref{sec:critical}.

Let us also shortly dwell, here, on some of the features of the so-called detailed balance (DB) dynamics \cite{CMW11,ELS1990}. We say that the spins located at one boundary of the lattice are (locally) in \textit{equilibrium} with the nearest reservoir if the rate $c^\textrm{DB}$ at which a spin value $\sigma_{i}'$ is introduced at the boundary site $i$ obeys the following DB condition:
\be
\label{db}
% %c^\textrm{DB}(\sigma,\sigma^i)
 c^\textrm{DB}(\sigma_i')
% = \textrm{min}\left[1,\left(\frac{1+m_\textrm{res}\sigma_i'}{2}\right)e^{-\beta \Delta H}\right], \qquad \Delta H = H(\sigma^{i}) - H(\sigma),
= \textrm{min}\left[1,\left(\frac{1+m_\textrm{res}\sigma_i'}{1+m_\textrm{res}\sigma_i}\right)e^{-\beta \Delta H}\right], \qquad \Delta H = H(\sigma'|\sigma^o) - H(\sigma|\sigma^o),
\ee
where $\sigma'$ denotes the configuration obtained from $\sigma$ by updating the spin $\sigma_i$ to $\sigma_i'$, and the term $(1+m_\textrm{res}\sigma_i')/2$ corresponds to the probability of drawing at random the spin $\sigma_i'$ from the reservoir with magnetization $m_\textrm{res}$.
We may then use an identity that follows from Eq.\ \ref{defbetah},
\be
\label{mh}
\frac{1+m_\textrm{res}\sigma_i'}{1+m_\textrm{res}\sigma_i}=e^{\betah_\textrm{res}(\sigma_i'-\sigma_i)},
\qquad \forall \sigma_i,\sigma_i'\in\{-1,+1\}
\ee
to rewrite \eqref{db} more conveniently as
\be
\label{db2}
 c^\textrm{DB}(\sigma_i')
=  \textrm{min}\left(1,e^{-\beta \Delta \mathcal{H}}\right), \qquad
\Delta \mathcal{H} = \mathcal{H}(\sigma'|\sigma^o) - \mathcal{H}(\sigma|\sigma^o)
\ee
where
\be
 \mathcal{H}(\sigma|\sigma^o)= H(\sigma|\sigma^o)-  \sum_{i\in(x_b,y)} h_\textrm{res} \sigma_i
\ee
corresponds to the hamiltonian of an Ising model with opposing magnetic fields $h_\textrm{res}$ acting on
all boundary spins.

\subsection{Observables}
\label{sec:obs}
The presence of the two reservoir at the boundaries of the domain $\Lambda$ with different magnetizations induces a magnetization flow  across the  system.
The magnetization current along a given horizontal bond is measured by counting the number of positive spins that cross this bond from left to right minus the number of those crossing this bond in the opposite direction, and dividing this number by time. The stationary current $J$ then corresponds to the long time limit of this rate \cite{CGGV18}, averaged over all bonds, or alternatively, averaged over boundary bonds only, provided a stationary limit exists.

\begin{table}
\centering
\begin{tabular}{lll}
\hline\hline
Symbol & Definition & Description \\
\hline
$m_+$ & Parameter & Magnetization of reservoir ${\cal R}_+$ \\
$m_-$ & $=-m_+$ & Magnetization of reservoir ${\cal R}_-$\\
$m_\textrm{res}$ & res\ $\in\{-1,+1\}$ & stands for $m_+$ or $m_-$ \\
$m_\beta$ & \eqref{mbeta} & Onsager's equilibrium magnetization\\
$\overline{m}_x$ & \eqref{defmeanmx} & Mean magnetization at position $x\in \{1,..,L\}$\\
$\overline{m}_L$ & $=\overline{m}_{x=L}$ & Mean magnetization at the right boundary \\
$\overline{m}_L^\textrm{X}$ &  & $\overline{m}_L$ obtained using model X \\
$\meq$ & $=\overline{m}_L$ & Mean absolute magnetization at the right boundary for ${\cal E}$-model\\
$\meq^\textrm{X}$& $=\overline{m}_L^\textrm{X}$ & $\meq$ obtained using model X $\in\mathcal{E}$ \\
$\mcrit$ & & Largest $m_+$ value for which $J=0$\\
$\mcrit^\textrm{X}$ & & $\mcrit$ obtained using model X $\in\mathcal{N}$ \\
$m_\textrm{bump}$ & & maximum $\overline{m}_x$ value close to boundary of the bump profile\\
$m_o$ & $=\overline{m}_L^\textrm{DB-L4}$ & for the case of $m_+=0$\\
$\overline{m}$ & $=L^{-1}\sum_{x=1}^L\overline{m}_x$ & Mean bulk magnetization\\
$m(r)$ & & Magnetization at reduced coordinate $r\in[0,1]$\\
$m_u$ & & stability region for bump $m\in[m_u,\mcrit]$ \\
%$m_+^\textrm{eff}$ & $= \overline{m}_L^\textrm{DB-L4}(\beta,m_+)$ & currently used for \figref{figtransformed} \\
$m_+^\textrm{eff}$ & \eqref{meff} & effective magnetization, $\overline{m}_L^\textrm{ISF-L4}(\beta,m_+^\textrm{eff}) = \overline{m}_L^\textrm{DB-L4}(\beta,m_+)$\\
\hline\hline
\end{tabular}
\caption{\label{tabnotation}Notation for magnetizations considered in this work. Whenever a symbol $h$ is used, it is related to the corresponding magnetization by $m=\tanh(\beta h)$, where $\beta$ denotes inverse temperature, c.f. Tab.\ \ref{Table1}.}
\end{table}

A typical stationary state configuration  can be surveyed by introducing the {\em stationary magnetization profile} $\overline{m}_x$, $x=1,\ldots,L$,  that is obtained as follows: we measure the average magnetization along the column $x$
\begin{equation}
m_x(t) =  \frac 1 L\sum_{y=1}^{L} \sigma_{(x,y)}(t), \qquad \forall x\in \{1,..,L\}
\label{defmx}
\end{equation}
at each MC step $t$, and then we take the average $\overline{m}_x$ over the set of collected values in the course of MC, with $T$ steps in total,
\begin{equation}\label{defmeanmx}
 \overline{m}_x = \frac{1}{T}\sum_{t=1}^T m_x(t) , \qquad \forall x\in \{1,..,L\}
\end{equation}
In practise, the expensive sums are never evaluated during the MC, as each spin or bond flip causes only changes of $m_x(t)$. The mean boundary magnetizations are $\overline{m}_1$ and $\overline{m}_L$. For the ${\cal E}$--models $\overline{m}_L$ is denoted by $\meq$. The notation used in this work is summarized in Tab.\ \ref{tabnotation}.

By combining reservoir mechanisms and b.c.s we have defined nine different models,
a set of three equilibrium models ${\cal E}\equiv\{$NSF-L4, NSF-FB, NSF-SB$\}$, a set containing the local equilibrium models ${\cal L}\equiv\{$DB-L4, DB-FB, DB-SB$\}$, and the corresponding three nonequilibrium models ${\cal N}\equiv\{$ISF-L4, ISF-FB, ISF-SB$\}$.
The investigation of the ${\cal E}$, ${\cal L}$, and ${\cal N}$ dynamics, equipped with the different b.c.s, will allow us to comment, in Section\ \ref{sec:results}, on the relation between the magnetizations $\mcrit$ and $\meq$, mentioned above, for different values of $\beta$.

\subsection{Implementation details}
\label{sec:implem}

All results to be presented are obtained for 2D grids with $L=40$. Each MC step corresponds to
a single attempted bond or boundary spin flip at a randomly selected site. If not otherwise mentioned,
(i) the initial configuration, referred to as ``{\em configuration A}'', has $\sigma_i=-1$ for $x\in\{1,..,20\}$
and $\sigma_i=+1$ for $x\in\{21,..,40\}$ (the configuration shown for $h_+=5.35$ in \figref{schematic-beta=1-ISF-L4} later below), and (ii) each simulation for given set of
parameters $\beta$, $m_+$ runs for $\tau=10^{13}/N_b$ steps. Contributions to observables like the current $J$ and
magnetization profiles $\overline{m}_x$ are computed at each MC step, as each bond of spin flip gives rise to an increment of decrement of these quantities as opposed to the total energy, which is for this reason only
calculated each $N_b$ MC steps.
To save computing time further all nine models ${\cal E}$, ${\cal L}$, and ${\cal N}$ are calculated during a single run, and share their random numbers. This way identical bonds or spins are selected for an attempted MC move simultaneously, and the cost for the calculation of acceptance rates is minimized. Exponentials $\exp(-\beta \Delta H)$ are nowhere calculated but looked up from its very limited set of possible values.
Relevant neighbors for the calculation of $\Delta H$ are saved once for each bond and each model, which completely eliminates the calculation of b.c.s such as L4 or periodic b.c.s in $y$--direction. We have calculated the current $J$ independently from spins at the left and right boundary. For all stationary results to be presented, they are identical within statistical errors.

To capture the details as a function of $m_+$,
we plot results versus the dimensionless quantity
$\betah_+\equiv \tanh^{-1}(m_+)$. This representation is advantageous to blow up the region of $m_+$ close to 1, where interesting phenomena appear in a very narrow interval of $m_+$ \cite{CGGV18}.
To ensure that measurement points are equidistantly spaced in this representation we
choose $\betah_+$ equidistantly spaced between $0$ and $6.5$, at a resolution of $0.05$. In addition,
we calculate the result for $m_+=1$ (corresponding to $\betah_+=\infty$)
and eventually add it to plots as a filled marker. The 2D maps are obtained at a resolution of $\Delta \beta=0.01$ and $\Delta(\beta h_+)=0.05$. A function like $\mcrit(\beta)$ is extracted from the $J$--map
as contour line at $J=0$. The magnetization $m^\textrm{eff}_+$ is extracted by first inverting the $\overline{m}_L^\textrm{ISF-L4}$ map, using a continuous interpolant. The quantities $m^\textrm{eff}_+$  and $\overline{m}_L^\textrm{ISF-L4}$ are introduced in Sec.\ref{sec:critical}, see also Tab.\ref{tabnotation}.

\section{Results and discussion}
\label{sec:results}
In this section we present the results of our numerical  simulations of  the nonequilibrium steady state (NESS) attained by the ${\cal N}$--models  ISF-L4, ISF-FB, and ISF-SB introduced in Section\ \ref{sec:model}. Moreover, when convenient, we shall also compare the results obtained with the same b.c.s in presence of DB dynamics, i.e., ${\cal L}$--models DB-L4, DB-FB, and DB-SB. The equilibrium ${\cal E}$--models NSF-L4, NSF-FB and NSF-SB will also be considered for the purpose of comparison.  We aim at elucidating the nature of the NESS and the role played by the details of the b.c.s and reservoir mechanisms.  In particular, we will  focus on the stationary magnetization profiles $\overline{m}_x$ and currents $J$, defined in Section\ \ref{sec:obs}. In all the simulations to be considered below, the initial datum is the ``configuration A'' introduced in Section\ \ref{sec:implem}, and whose total magnetization is zero.

Since our models depend on  $L$, $\beta$ and $m_+$, the corresponding NESS will  also depend on the same set of parameters in a manner which is far from obvious. It is therefore of crucial importance to choose properly these quantities in our simulations.
The dependence of the NESS on the inverse temperature $\beta$ is likely to be dictated by the presence of the ferromagnetic phase transition in the equilibrium system. Thus, different behaviors are to be expected in the two temperature regimes:  standard phenomena  (e.g. Fick's law)  at high temperatures and a more complex and intriguing picture in the low temperature phase.  As representative instances of the two different regimes, we have chosen  $\beta=0.3$ and $\beta=1$, respectively lower and higher than the critical value of  $\beta_c$ in \eqref{betac}.

We recall that the linear size of the lattice
$\Lambda$ in our simulations, for which results are presented, is chosen as $L=40$, as in \cite{CGGV18}. Indeed, to the best of our knowledge, this lattice appears sufficiently large to suppress the most evident finite-volume effects that can be observed  for $L$ ranging between 10 and 40, while being sufficiently small to allow for a detailed numerical study within an acceptable computational workload.   However, we cannot exclude  the existence of significant finite-size effects for  systems with sizes  of the order of our size $L=40$. These issues will be discussed further in what follows.

The main results of  \cite{CGGV18}, that will be extended and deepened here, concern the behavior of the stationary current $J$ of the ISF-L4 model, as the magnetization $m_+$  ($m_-$) of the reservoir(s) is varied. We recall them briefly here.
\begin{enumerate}
\item As the value $m_+$ is decreased from its maximum value 1, the flux $J(m_+)$ is first negative and, crossing a critical value  $\mcrit$, it becomes positive. More precisely, for $m_+>\mcrit$ the current is negative, flowing from the reservoir with positive magnetization $\mathcal R_+$ to the one with negative magnetization  $\mathcal R_{-}$ (in agreement with the  Fick's law).   For $m_+<\mcrit$ the current is reversed, flowing against the magnetization gradient, that is from $\mathcal R_{-}$ to  $\mathcal R_+$ (in violation of the Fick's law).  In the latter case we have the phenomenon called  {\em uphill diffusion}.

\bigskip
\item The flip of the current in passing from downhill to uphill diffusion is connected to a change in the structure of the NESS \cite{CGGV18}. This can be detected by studying the stationary magnetization profile $\overline{m}_{x}$ or by inspection of the typical spin configuration of the NESS.
In  \cite{CGGV18} two typical magnetization profiles have been identified: {\em instanton} and {\em bump}.

\medskip
The {\em instanton} profile corresponds, in the spin configurations, to the presence  of an interface separating two plus and minus equally sized phases close to  $\mathcal R_+$ and  $\mathcal R_-$ respectively, that is with the interface  placed in the middle of the lattice (cf. also \cite{SOS16} for an interpretation of the interface based on nonequilibrium thermodynamics).   A sketch of the instanton profile, in a continuous limit representation $m(r)$, where $r=(x-1)/(L-1)$, is given in \figref{schematic-profile}. The figure shows that, for $r$ moving from the right reservoir  $\mathcal R_+$ ($r=1$) to the center of the lattice ($r=1/2$), the profile $m(r)$ decreases from  $m_+$ (the magnetization imposed by $\mathcal R_+$) to a value which is expected to be the equilibrium spontaneous magnetization $m_\beta$ given in \eqref{mbeta}. At the center of the system, $m=1/2$, the magnetization jumps down to $-m_\beta$ and then keeps decreasing up to the left reservoir ($r=0$), where it takes the value $m_-$. Obviously, at finite volumes the sharp jump is replaced by a layer, i.e. a narrow region of sites in which the profile changes rapidly. Examples of steady magnetization profiles with an instanton-like shape are shown in the panel with $\betah_+=5.35$  of \figref{fig1.d1} for the ISF-L4, ISF-FS, ISF-SB models. The corresponding  spin configuration, for the ISF-L4 case, is shown in \figref{schematic-beta=1-ISF-L4}, where it is also evident that $J$ is negative in this state.

\begin{figure}[tb]
\centering
\includegraphics[width=0.6\textwidth]{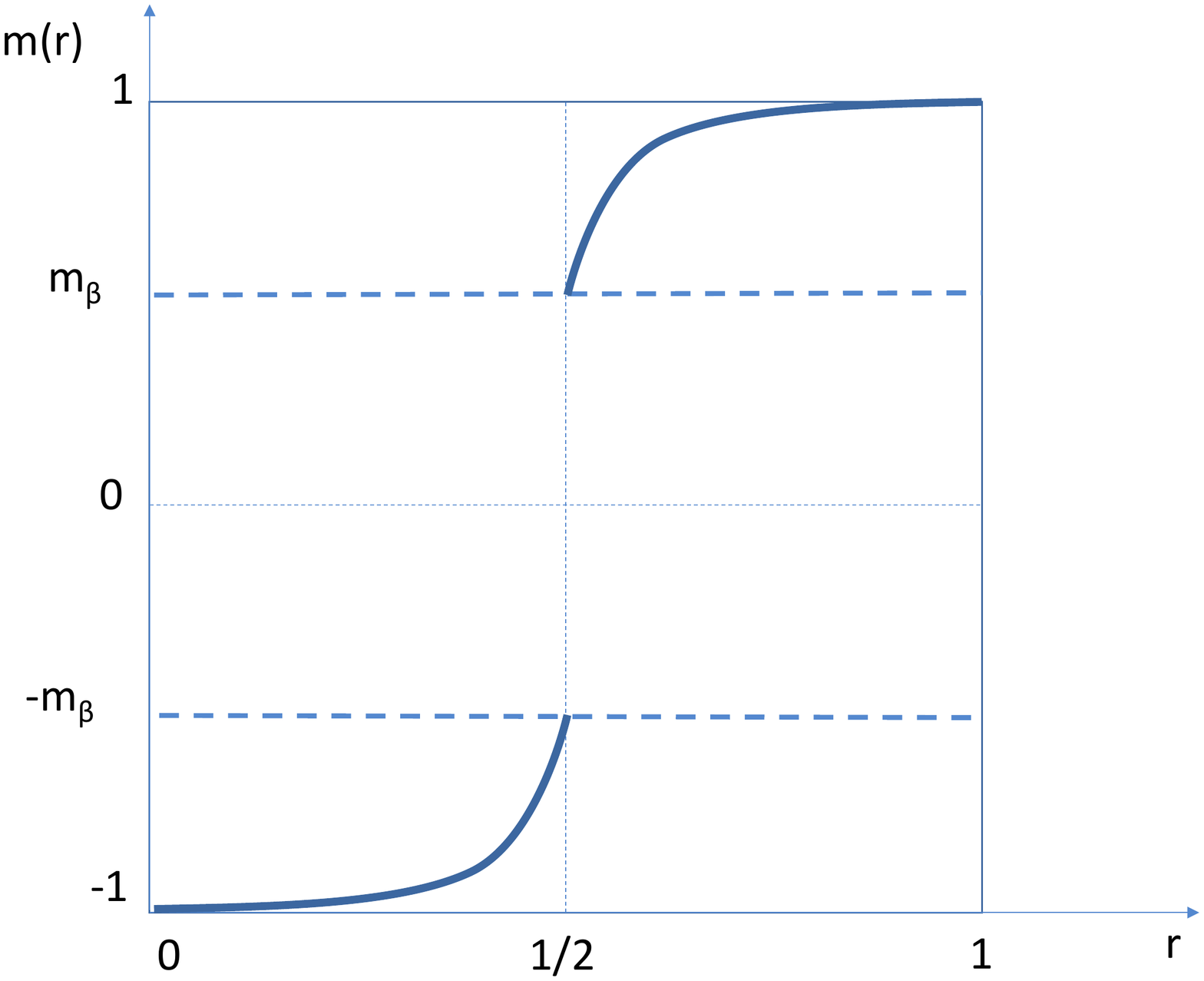}
\caption{\label{schematic-profile}Sketch of the magnetization profile $m(r)$ in the macroscopic coordinate $r=(x-1)/(L-1)\in[0,1]$, when $m_+=1$ and $\beta>\beta_c$.
For $\beta<\beta_c$, $m_{\beta=0}=0$, the gap is absent, and the profile becomes linear at $\beta=0$.}
\end{figure}

\begin{figure}[tb]
\centering
\includegraphics[width=0.8\textwidth]{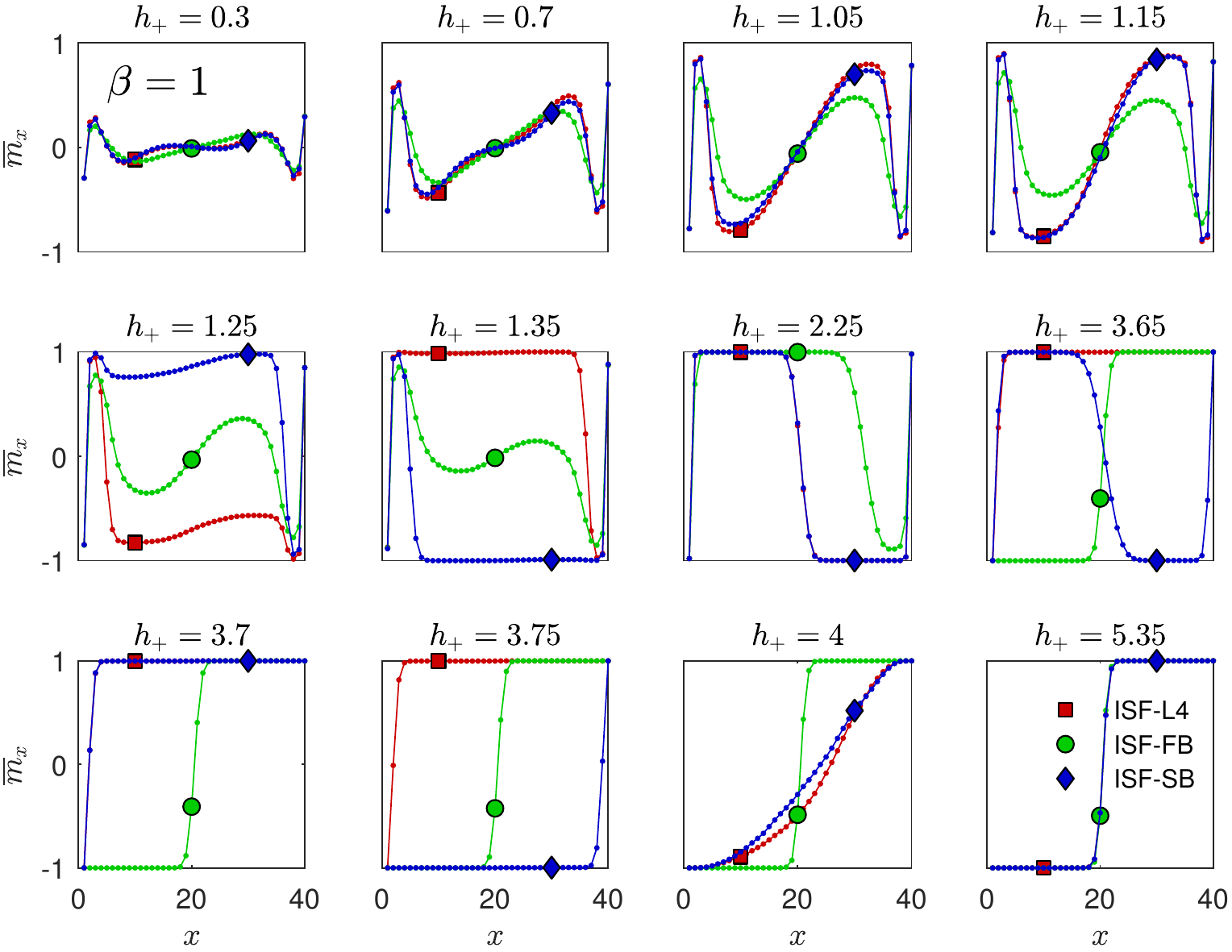}
\caption{\label{fig1.d1}Stationary magnetization profiles $\overline{m}_x$ versus $x$ for the three nonequilibrium models at $\beta=1$, i.e. $\beta \gg \beta_c$, a temperature well below the critical temperature
for the corresponding equilibrium system.
$40\times 40$ grid. $\tau=10^{13}/N_b$ for each $h_+$ value. Note that $h_+=\beta h_+$ for the present case.
}
\end{figure}

\begin{figure}
\centering
\includegraphics[width=0.75\textwidth]{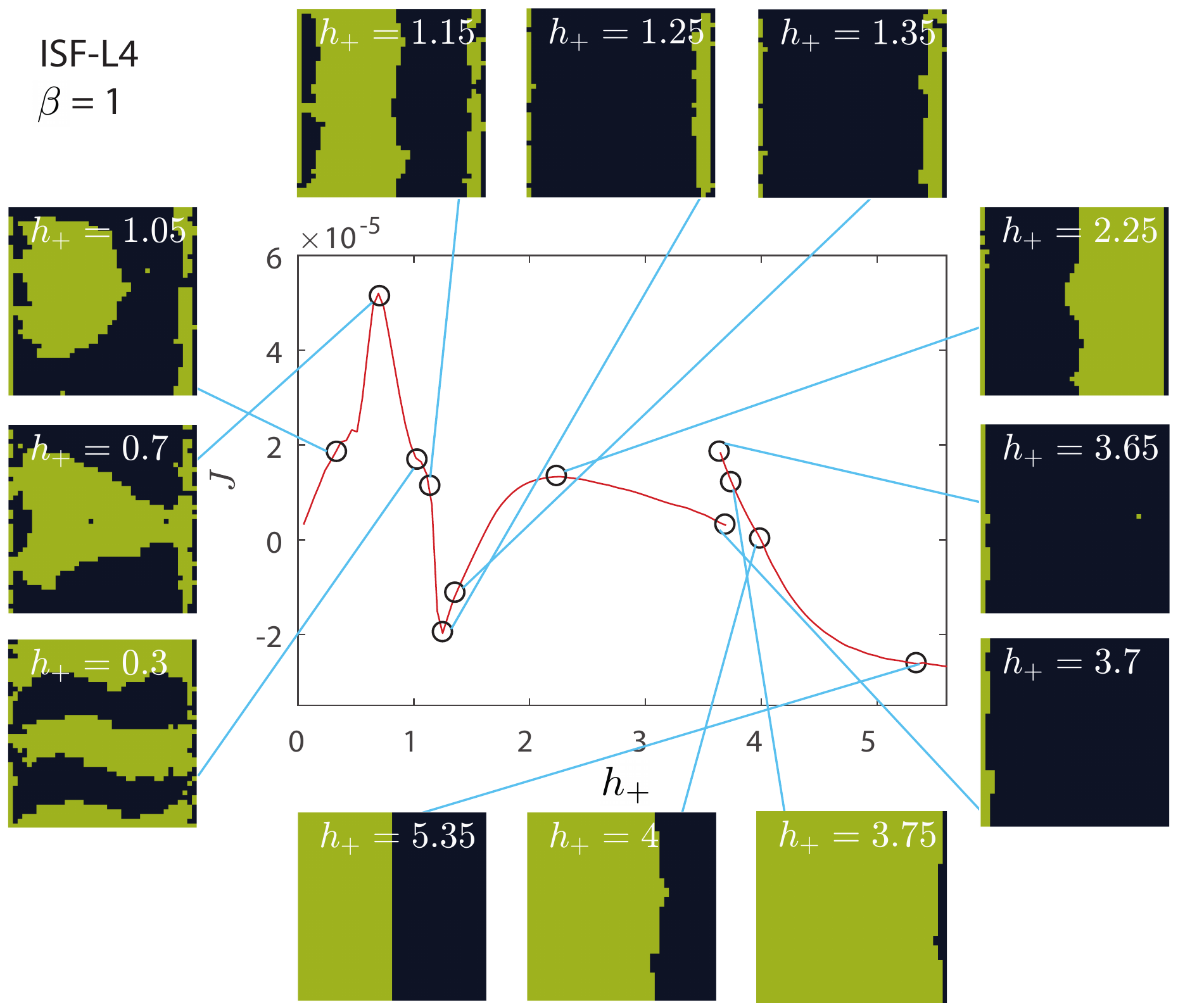}
\caption{\label{schematic-beta=1-ISF-L4}Nonequilibrium current $J$ versus $h_+$ for $\beta=1$ (ISF-L4), together with
selected configurations at time $t=\tau=10^{13}/N_b$.
The corresponding magnetization profiles
are shown in \figref{fig1.d1} in concert with profiles for the other two models (ISF-FB and ISF-SB).
The metastable region at $h_+\approx 3.7$ ($m_+\approx 0.9988$) is investigated in more detail below. The current changes three times the sign.
Between panels marked by $h_+=1.15$ and $1.26$, between $h_+=1.36$ and $2.24$, as well as between $h_+=3.75$ and $4.0$.
It reaches its maximum value at about $h_+=0.71$. Color code: spin +1 (black), spin -1 (olive). $40\times 40$ grid.
}
\end{figure}

\medskip
The  {\em bump} profile occurs when the typical spin configuration of the NESS presents an interface close to one of the vertical boundaries.  In this case a sea of (say) positive spins emanating from $\mathcal R_+$ invades the most part of the lattice $\Lambda$, leaving a small strip adjacent to $\mathcal R_-$ for the negative spins. The profile of $\overline{m}_{x}$ for $x$ ranging from the right ($x=L$) to the left ($x=1)$ boundary, increases steadily from $m_+$ up to a maximum value $m_\textrm{bump}>m_+$, which is reached for $x$ close to 1. Then $\overline{m}_{x}$ jumps suddenly from $m_\textrm{bump}$ down to $m_-$ at $x=1$, forming a boundary layer close to the left boundary. A similar profile is obtained by changing in the previous description the sign of magnetization and exchanging  left with right. In any case, in the presence of a bump the current flows in the `wrong' direction, producing {\em uphill diffusion}. The bump appears, for instance, in the  ISF-L4 model for $\betah_+=3.75$,  see \figref{fig1.d1}. In \figref{schematic-beta=1-ISF-L4} it is shown, for the same value of $\betah_+$, the spin configuration with the boundary layer close to $\mathcal R_+$  and the evidence for the uphill current, i.e.  $J>0$.

\bigskip
\item The instanton profile, that sustains a negative current, is stable for $m_+\in(\mcrit,1)$.  This means that,  perturbing the \textit{configuration A}  initial datum or even taking a random initial condition  for $m_+>\mcrit$, the dynamics leads asymptotically to the instanton profile for $\overline{m}_{x}$. This is the {\em stable phase}. The situation is quite different and much more intricate for $m_+\in (0, \mcrit)$. Here the instanton loses stability and we have numerical evidence that, at least for not  too low $m_+$ values, the bump profile is stable.  Thus, there should exist a further critical value
$m_{u} <\mcrit$ that defines the stability region for the bump $(m_u, \mcrit)$. The nature of $m_u$ is not well understood, but it is conjectured that its presence should be a finite-volume effect, since in the large volume limit the interval $(m_u, \mcrit)$ is supposed to shrink to the empty set. Indeed, some heuristic reasoning leads to  the  estimate $O(L^{-2/3})$ for its length. This phase has been termed {\em metastable} in \cite{CGGV18}. Below $m_{u}$ the picture is much more complex and less studied. In the next section we will show that decreasing further $m_+$ towards 0, the system undergoes an intricate sequence of  different stationary states. This is the {\em unstable phase}, that was also referred to as the  ``weakly unstable'' or  ``chaotic region'' in \cite{CGGV18}.

\end{enumerate}

\subsection{Behavior of $J$ in external magnetization-temperature space}

\begin{figure*}
\centering
\includegraphics[width=6cm]{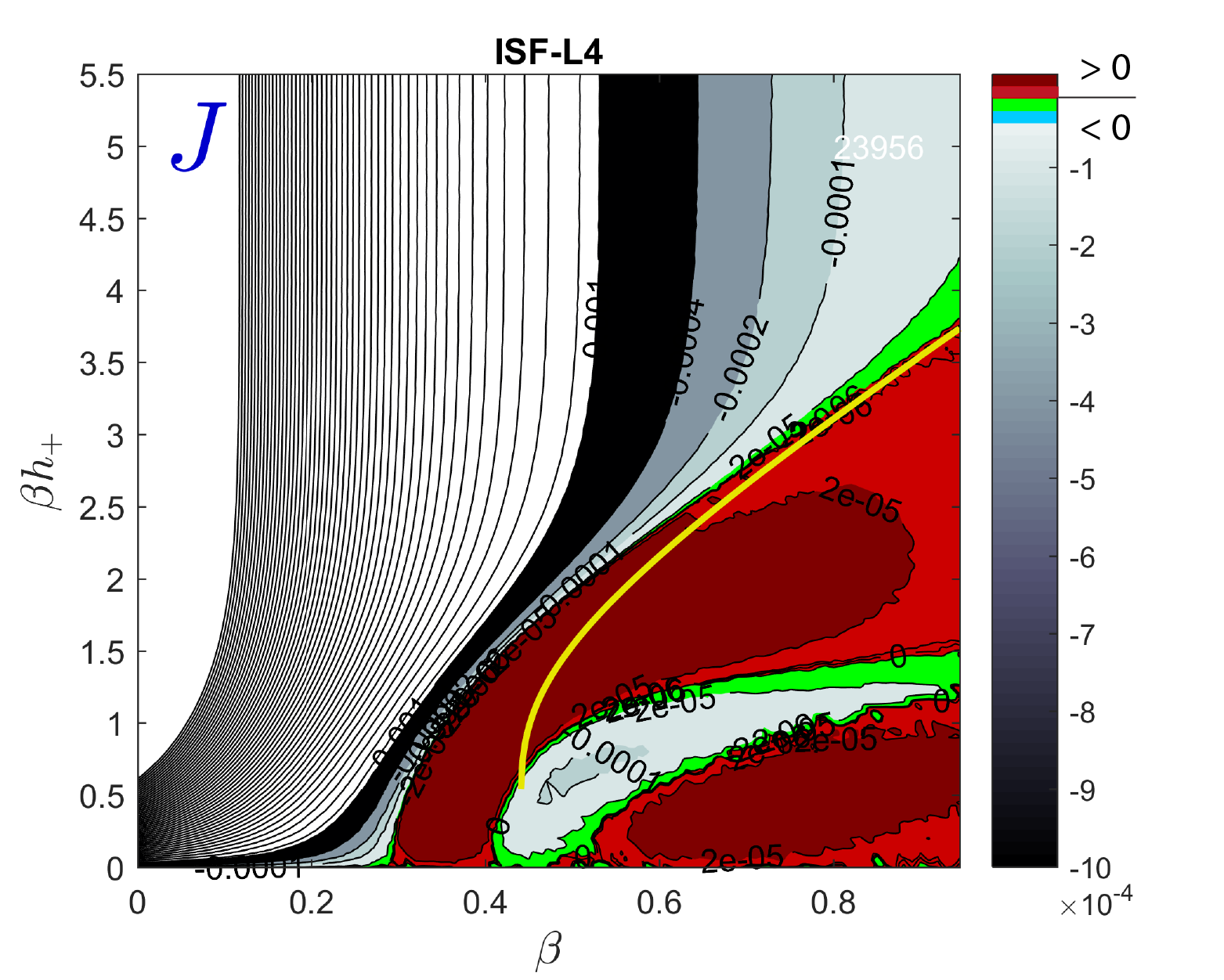}
\includegraphics[width=6cm]{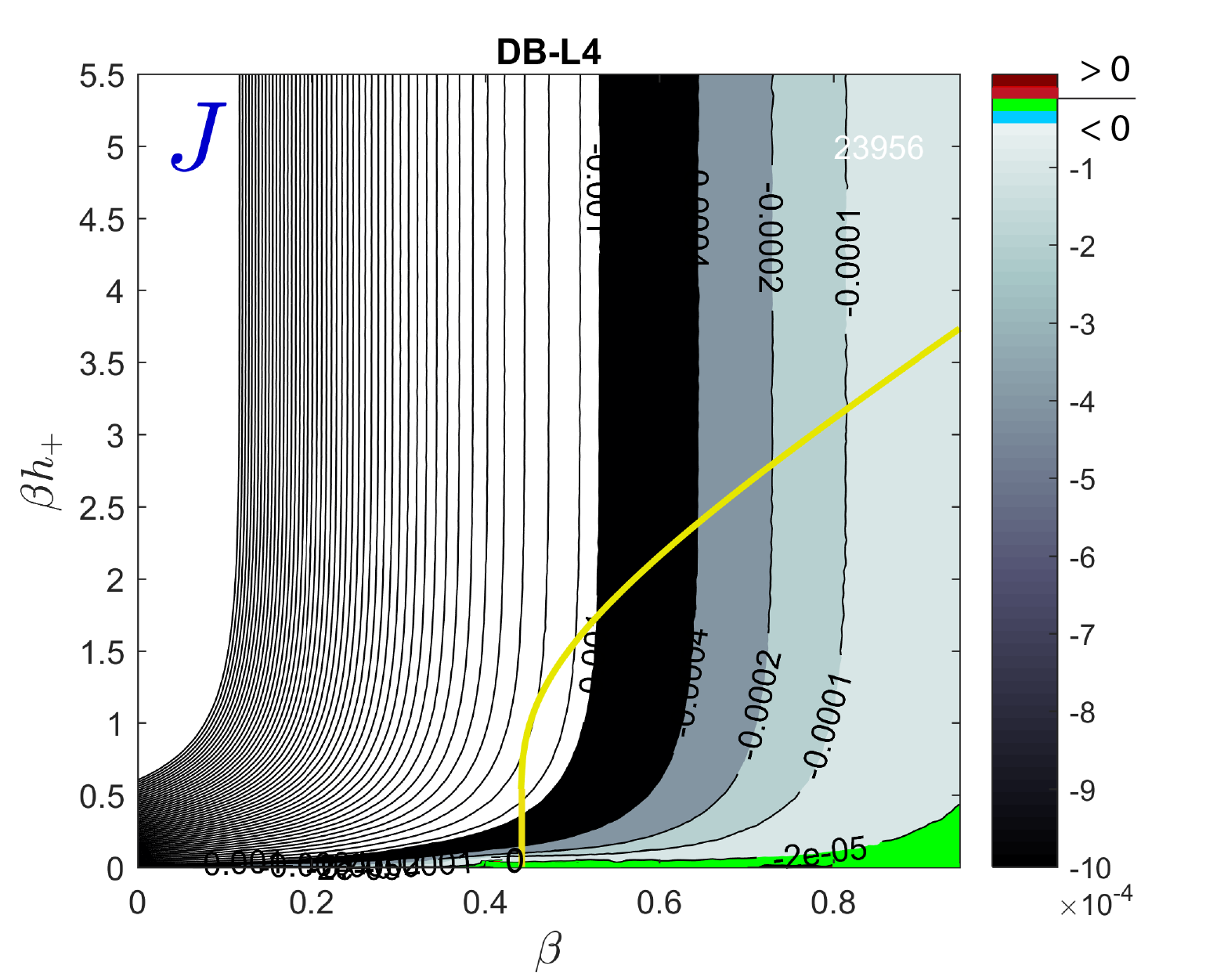}\\
\includegraphics[width=6cm]{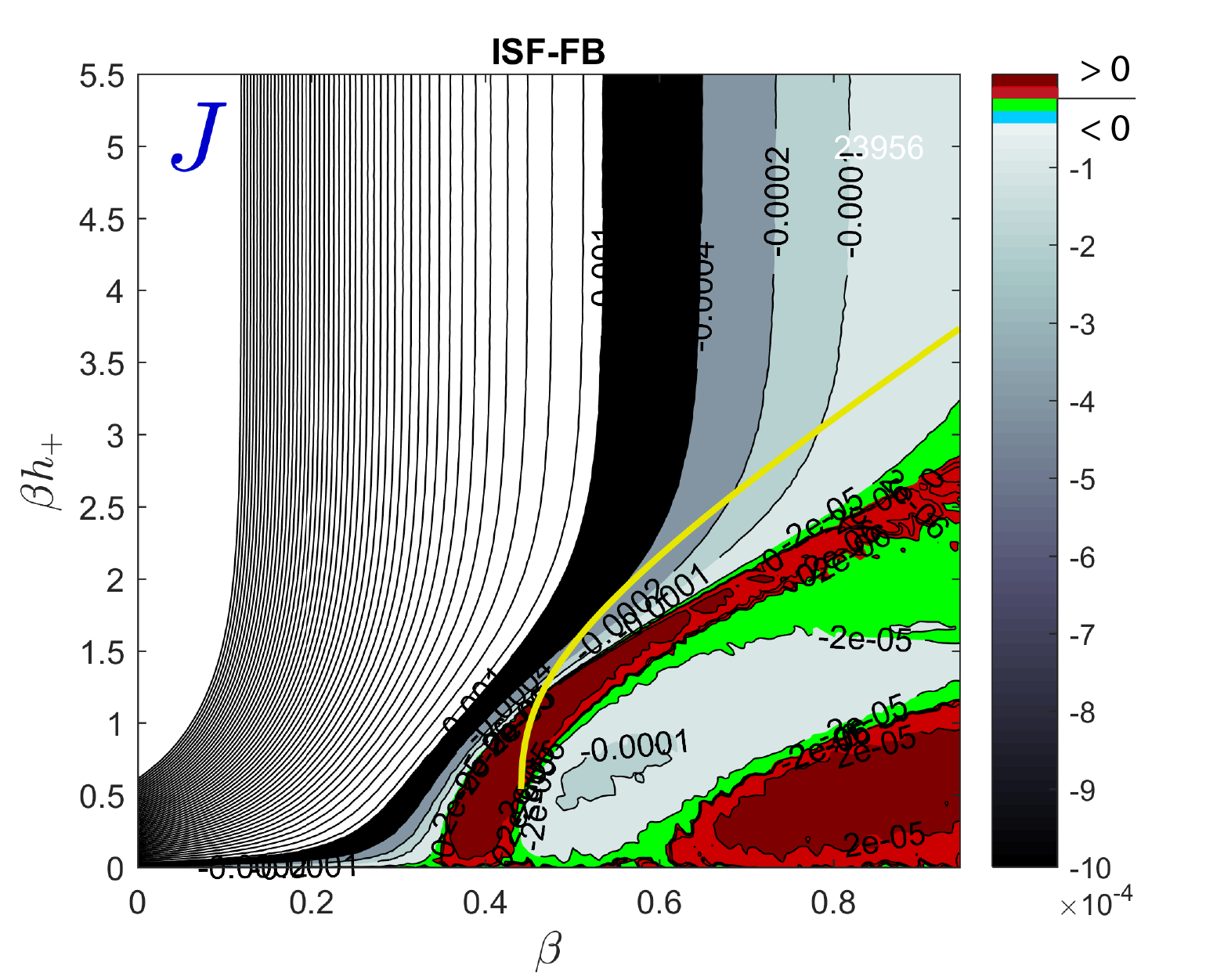}
\includegraphics[width=6cm]{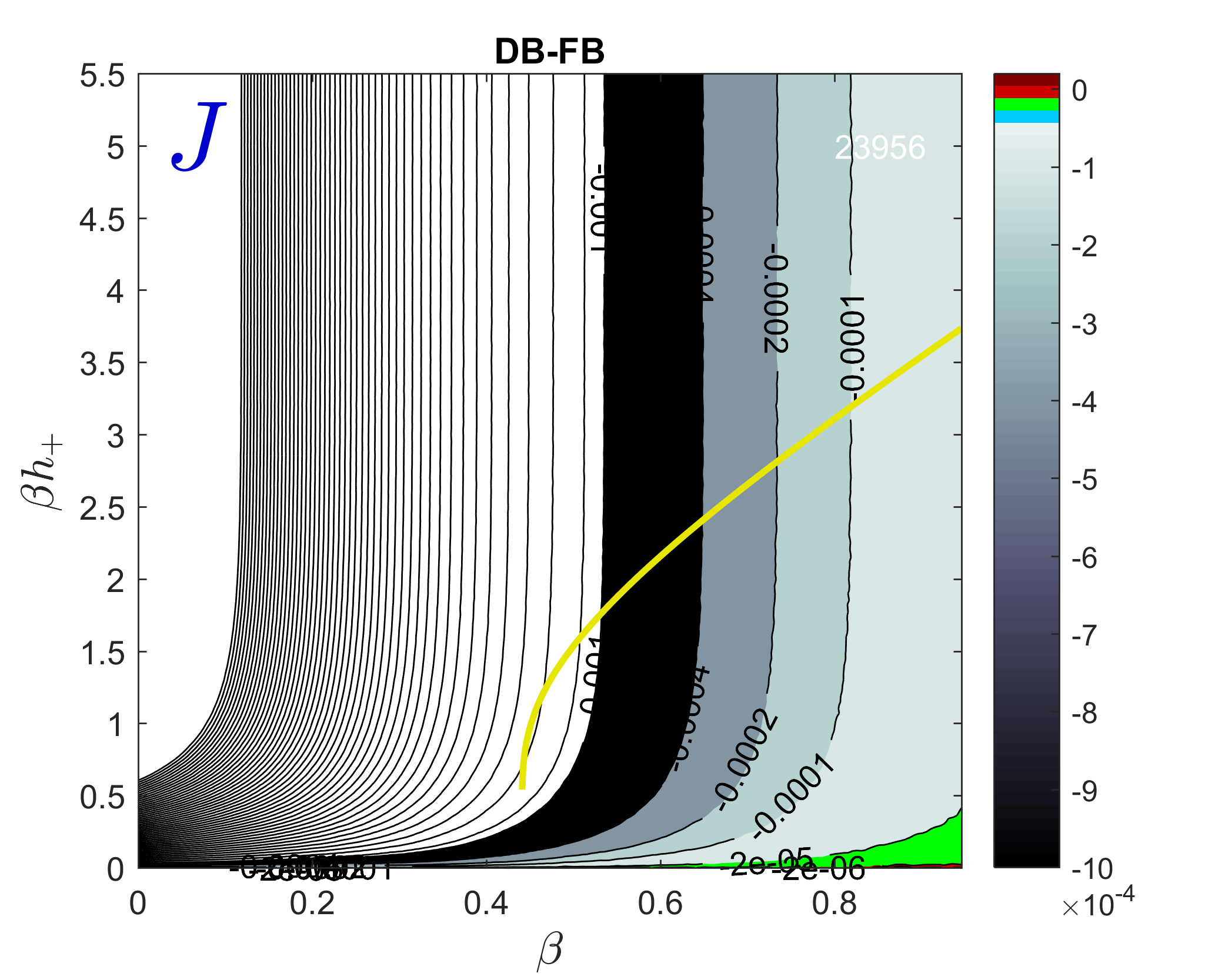}\\
\includegraphics[width=6cm]{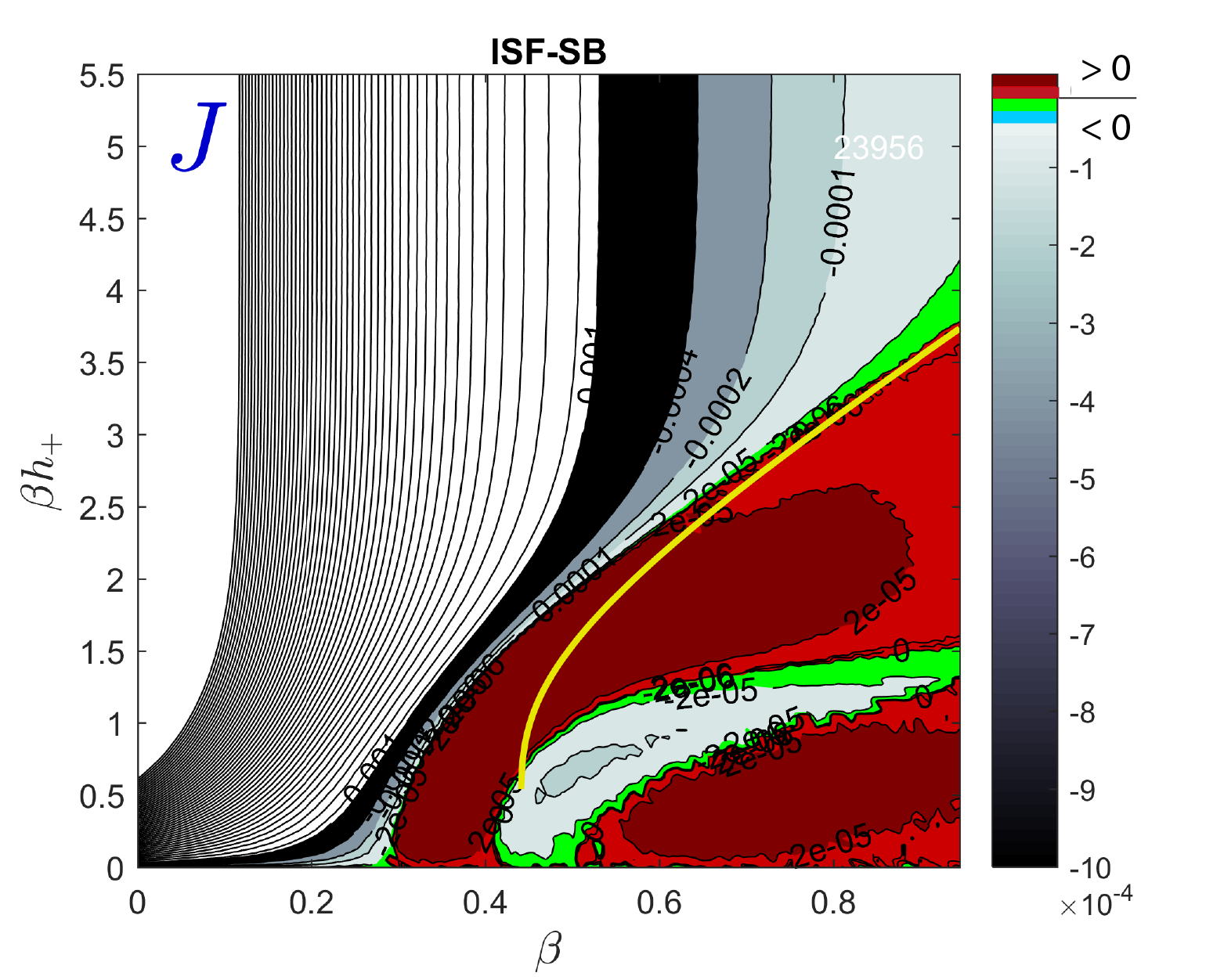}
\includegraphics[width=6cm]{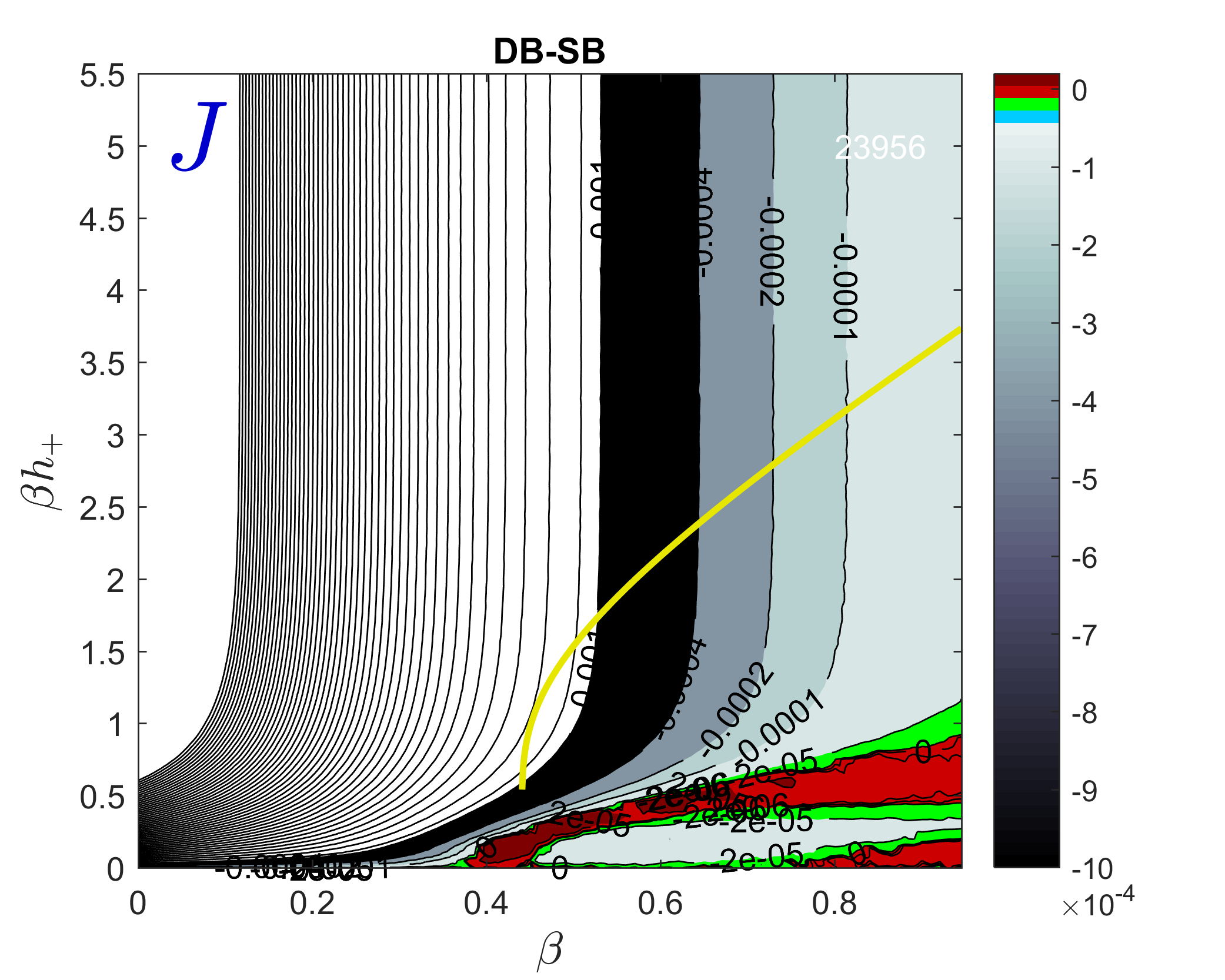}
\caption{2D maps showing the values of $J$ as a function of $\beta$ (horizontal) and $\betah_+$ (vertical) for the three ${\cal N}$--models (left column) and ${\cal L}$--models (right column). For the $\mathcal{E}$--models $J$ vanishes by definition. Regions with positive $J$ are colored in red, adjacent regions with negative $J$ are colored green, remaining regions with $J<0$ use a grayscale; $J=0$ at $\betah_+=0$, i.e., $m_+=0$. As is evident from these 2D maps, and depending on the pathway in $\beta$--$\betah_+$--space, $J$ changes several times its sign for all but the DB-L4 and DB-FB models. The yellow line had been added as reference, it corresponds to Onsager's exact result for the 2D Ising model, $\betah_{\beta}=\tanh^{-1}(m_\beta)$ with $m_\beta$ from \eqref{mbeta}. Height profiles of the three ISF-maps at two vertical lines, at $\beta=0.3$ and $\beta=1$, are shown in better resolution in \figref{fig1.e}. The uppermost contour line $J=0$ (borderline between red and green) for all ISF-models is shown in \figref{fig3.1}.
}
\label{maps}
\label{Jmap}
\end{figure*}

We shall discuss here  the 2D maps for the current $J$ in the parameter space spanned by $ \betah_+=\tanh^{-1}(m_+)$ and $\beta$ for the 3 different models ISF-4, ISF-FB and ISF-SB. The MC results are portrayed in \figref{Jmap} (left column). Remarkably, the ISF-L4 and ISF-SB dynamics seems to yield similar results for the current, while the FB b.c.s gives rise to a different scenario, in which the onset of uphill currents is somehow hindered. It should also be noted that, with the L4 and SB b.c.s, the Onsager's result for $m_\beta$, denoted by the yellow line in \figref{Jmap} is close, for large enough values of $\beta$, to $\mcrit$ since it marks closely the transition between regions with positive currents (colored in red, in the figure) and other regions with negative currents (colored in green). For smaller values of $\beta$, the yellow line no longer stays at the border between red and green regions, hence $m_{\beta}$ and $\mcrit$ substantially differ from one another, as it will also be seen in \figref{fig3.1}.

In Section\ \ref{sec:critical} we shall focus, in particular, on the behavior of the current as a function of $\betah_+$ along the vertical lines at $\beta=0.3$ and $\beta=1$.
An inspection of \figref{Jmap} reveals that positive 'uphill' currents $J>0$ (colored in red) are observed, with the ISF mechanism, when the parameters $\beta$ and $m_+$ of our model are suitably tuned. In particular, the parameter $\beta$ is required to be larger than a certain critical value $\beta_\textrm{crit}$. Note that the latter, due to finite size effects, might even significantly differ from the critical value $\beta_c$ in \eqref{betac} determined by Onsager (pertaining to the infinite volume Ising model).
The ISF mechanism is an essential ingredient to observe uphill currents.
It is worth comparing the ISF results shown in \figref{Jmap} (left column) with the corresponding results obtained with the DB dynamics equipped with the L4, FB and SB b.c.'s, see \figref{Jmap} (right column). An inspection of \figref{Jmap} confirms that while the L4 and FB b.c.'s give rise to a purely Fickian behavior, the DB dynamics equipped with the SB b.c. defined in \eqref{ghsp} yields uphill currents. The onset of positive currents even in presence of DB dynamics stems from the choice of the aforementioned boundary condition, which, as it stands, does not match properly with the DB dynamics. This aspect will be clarified in Section \ref{sec:mL}.

\subsection{Critical value $\mcrit$, NSF equilibrium magnetization $\meq$ and DB equilibrium magnetization $m_o$  }
\label{sec:critical}

The critical value $\mcrit$ separating the stable and the metastable phases is the threshold at which the current changes sign, i.e. where the current vanishes, $J(\mcrit)=0$.
As already said in the Introduction, in  \cite{CGGV18} it has been conjectured that in the large volume limit $\mcrit$ should  approach the equilibrium magnetization $m_\beta$. However,
in order to take into account the finite size effects, in \cite{CGGV18} it is also claimed that $\mcrit$ can be measured by evaluating the magnetization $\meq$ of the rightmost column of $\Lambda$
{\em in the absence of the reservoirs}, i.e. at equilibrium (NSF models).  The choice of the L4 b.c.s should entail that in the large volume limit $\meq$ approaches $m_\beta$.
Some numerical evidence  of this fact is given in \cite{CGGV18}  for the ISF-L4 case, where $\meq$ is computed for $L$ ranging from 10 up to 40 and $\beta=1$.
Here, we study the dependence of $\meq$  and $\mcrit $ on the parameter $\beta$, and compare the ISF-L4 case  with ISF-FB and ISF-SB. We denote by $\mcrit^\textrm{X}$, with
X\ $\in {\cal N}$,
the critical magnetization computed according to the nonequilibirum model $X$. Analogously, $\meq^\textrm{X}$, with X\ $\in {\cal E}$ is the quantity $\overline{m}_L$ defined by
\eqref{defmeanmx} computed in the equilibrium  model X.

For later use, we introduce also a quantity, named $m_o$, which is defined as $\overline{m}_L^\textrm{DB-L4}$ at $h_+=0$. We remark that, while $\meq$
is defined for the equilibrium bulk system in which the sole conservative Kawasaki dynamics acts, in defining $m_o$ we retain, in the presence of local equilibrium, the action of the two reservoirs even though with the same magnetization $m_+=m_-=0$ (corresponding to $h_+=0$). In this situation there is no net flux across the system induced by the gradient of magnetization at the boundaries, and a NESS is not established.  Thus we argue that, at least for sufficiently large systems, the two systems (i.e. the Canonical NSF-L4 and the Grand Canonical DB-L4 with $h_+=0$) should be equivalent and then  $\meq$ and $m_o$ should agree.
\begin{figure}[tb]
\centering
(a)\includegraphics[width=0.4\textwidth]{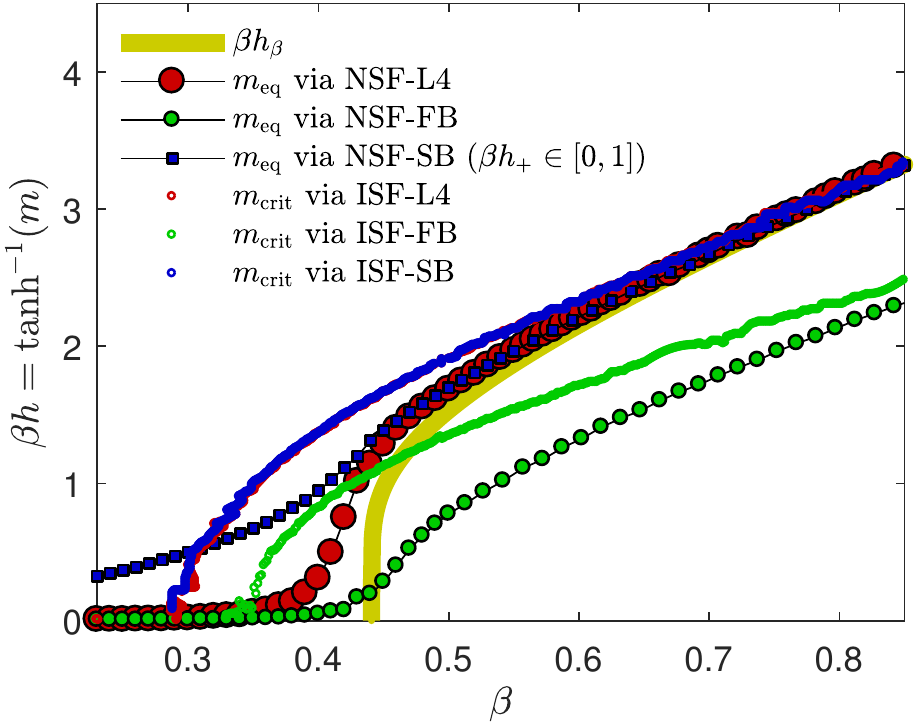}
(b)\includegraphics[width=0.4\textwidth]{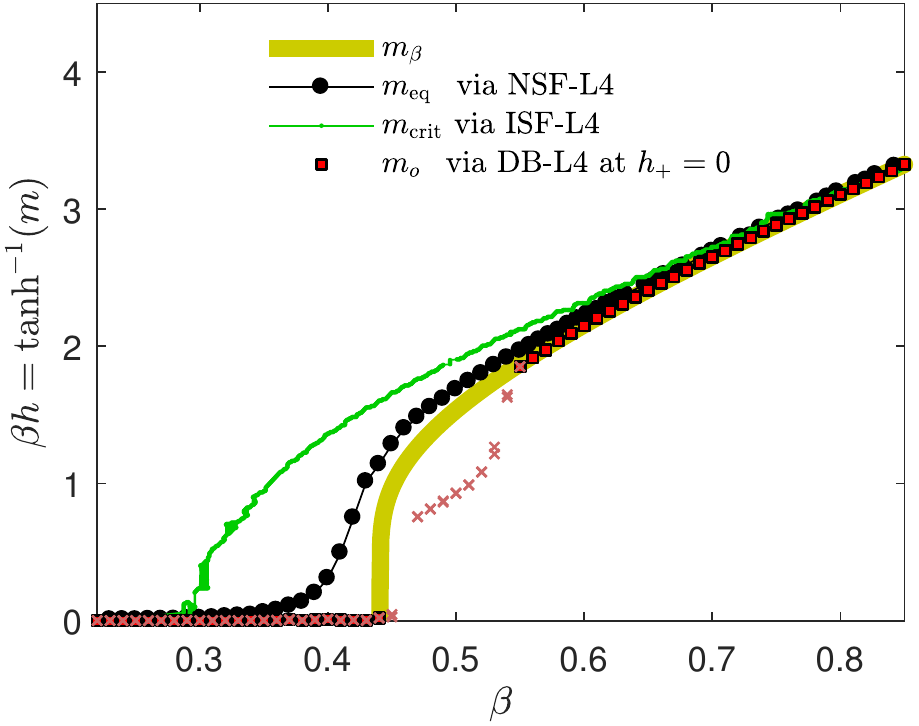}
\caption{\label{fig3.1} {\bf (a)}
Behavior of $\meq$, $\mcrit$, and $m_\beta$ (shown is $\tanh^{-1}(m)$ in each case) %$\beta\overline{h}_L=\tanh^{-1}(\meq)$, $\beta h_+^0=\tanh^{-1}(\mcrit)$ and $\beta %h_\beta=\tanh^{-1}(m_{\beta})$
as functions of $\beta$. Error bars are comparable with symbol sizes.
{\bf (b)} Behavior of $\meq$, $\mcrit$, $m_o$,
% $\beta\overline{h}_L$, $\beta h_+^0$ and $\tanh^{-1}(m_o)$,
all evaluated with L4 b.c., and $m_\beta$ as functions of $\beta$.
The data points for $m_o$ within the range $\beta\in[0.45,0.55]$ are marked by red crosses,
as they are not expected to match $m_\beta$. At these large temperatures the system frequently changes its
overall magnetization, and the mean magnetization at the boundary vanishes. The crosses represent the mean absolute magnetization at the boundary.
}
\end{figure}

In \figref{fig3.1} we show the critical vales $\mcrit^\textrm{X}$ for $\textrm{X}\in {\cal N}$,
the spontaneous magnetization  $m_\beta$, see \eqref{mbeta}, and  $\meq^\textrm{X}$, X\ $\in {\cal E}$ for $\beta \in (0, 1)$. $\mcrit^\textrm{X}$ is
the $m_+$--value closest to $1$ at which the current vanishes.
The figure shows that the status of the
conjecture regarding $\meq^\textrm{NSF-L4}$  and $\mcrit^\textrm{ISF-L4}$   in  \cite{CGGV18}  is  strongly  dependent on $\beta$. Indeed, for $\beta$ small  (say $	\beta<0.6$) the two quantities
are appreciably different one from the other and different from $m_\beta$. On the other hand,  $\meq^\textrm{NSF-L4}$ and $\mcrit^\textrm{ISF-L4}$ get closer and
closer to $m_\beta$ as $\beta$ is increased. For $\beta=1$, the value studied in \cite{CGGV18}, they almost coincide. The right panel of \figref{fig3.1} also shows the behavior of $m_o$, defined for the DB-L4 dynamics, as a function of $\beta$.

The role of b.c.s is also evident: for all $\beta$, the critical value $\mcrit^\textrm{ISF-FB}$ and $\meq^\textrm{NSF-FB}$ are clearly different from $\mcrit^\textrm{ISF-L4}$ and from $\meq^\textrm{NSF-L4}$ and $m_\beta$. Moreover, they seem to approach each other, as $\beta$ increases. It is also evident that ISF-L4 and ISF-SB coincide in the whole range of $\beta$. This is the first instance of a situation that we will meet frequently, i.e. the equivalence of ISF-L4 and ISF-SB models. Finally, we observe that all the quantities relative to models with b.c.s (i.e. L4 and SB), converge to $m_\beta$ as $\beta$ increases.

\subsection{Average magnetization at the boundaries}
\label{sec:mL}

\begin{figure}
\centering
\includegraphics[width=0.8\textwidth]{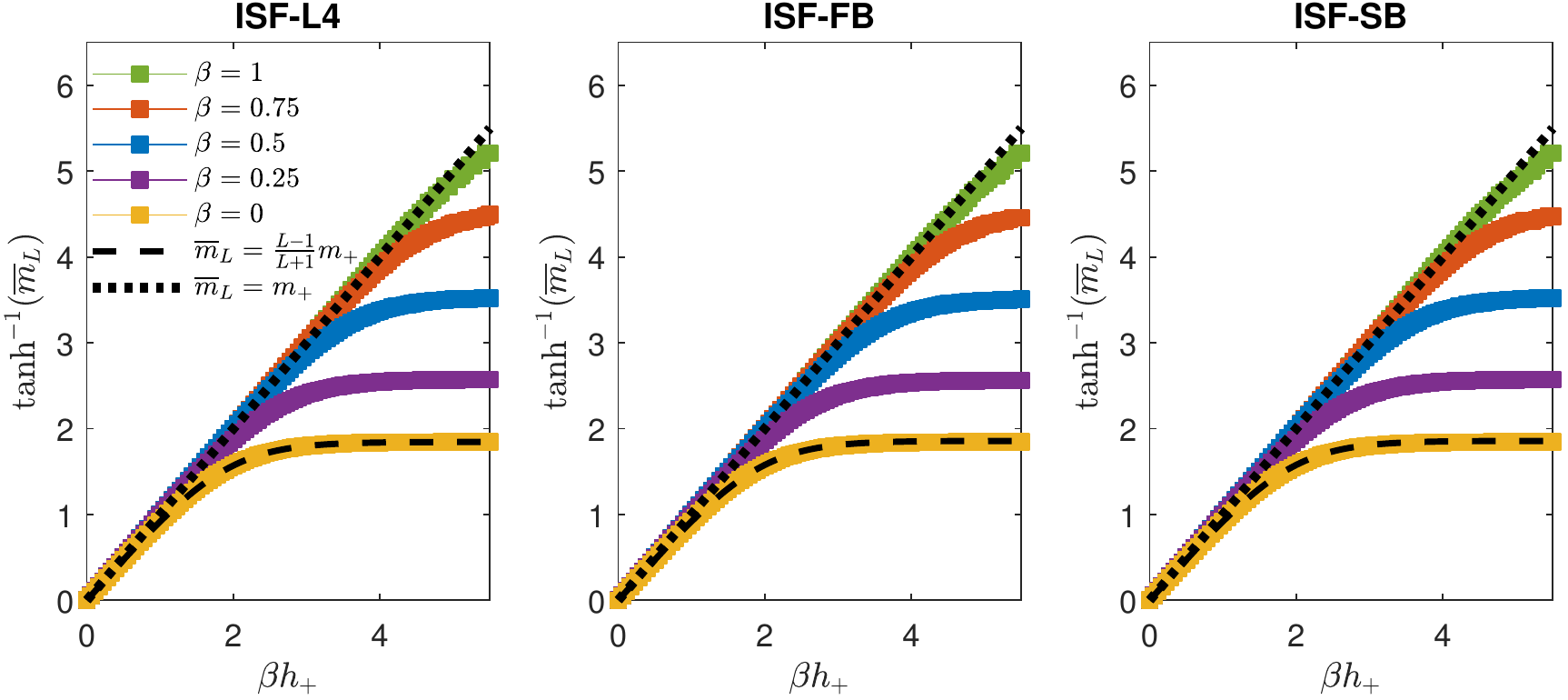}\\
\includegraphics[width=0.8\textwidth]{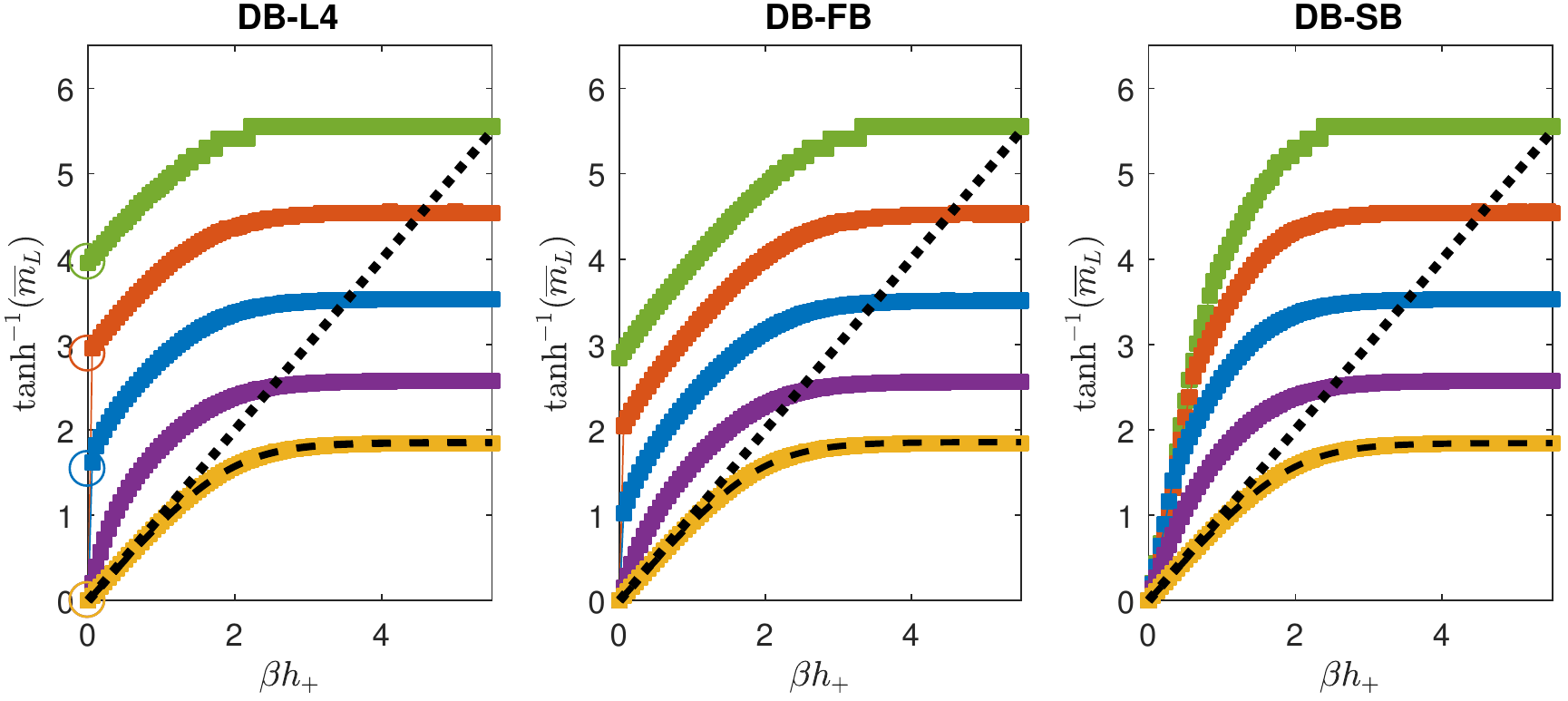}\
\caption{Behavior of $\tanh^{-1}(\overline{m}_L)$ as a function of the dimensionless magnetic field $\betah_+$, with $L=40$, for different values of $\beta$ and for the six $\mathcal{N}$-- and ${\cal L}$--models (from left to right): ISF-L4, ISF-FB, ISF-SB (top panels), and DB-L4, DB-FB, DB-SB (bottom panels). The dotted line corresponds to the curve $\overline{m}_L=m_+$, the dashed line represents the theoretical behavior of $\overline{m}_L$ for $\beta=0$, c.f. \eqref{beta0theory}. The large circles in panel DB-L4 at $\beta h_+=0$ (absence of external field) correspond to $\tanh^{-1}(m_\beta)$, cf. \eqref{mbeta}.}
\label{fig:mL}
\end{figure}

Another relevant feature of the ISF mechanism, outlined in \cite{CGGV18}, is that the stationary magnetization (averaged over the vertical direction) on the rightmost column, denoted as $\overline{m}_L$, see Tab.\ref{tabnotation}, attains a value close to $m_+$ when $\beta$ is large. \figureref{fig:mL} contains the result of MC simulations showing the behavior of $\overline{m}_L$ as a function of $\betah_+$ for different values of $\beta$, and for the various b.c.s considered above for the dynamics undergoing the ISF (top panes) and the DB (botom panels) updating mechanism. It is worth noticing that, for fixed $\beta$ and $\betah_+$, the b.c.'s seem not to affect significantly the resulting value of $\overline{m}_L$ with the ISF updating mechanism.
Another observation concerns the dependence of $\overline{m}_L$ on $\beta$, for a given $\betah_+$. \figureref{fig:mL} shows that, with the ISF mechanism (top panels) $\overline{m}_L$ attains a value that is closer and closer to $m_+$ when $\beta$ is large, e.g. at $\beta=1$. For small values of $\beta$ the two values $\overline{m}_L$ and $m_+$ start to deviate significantly, because then spin diffusion from the boundaries to the bulk of the lattice plays a major role.

It is also worth noticing that in \figref{fig:mL} the behavior of the DB-L4 and DB-SB models differ from one another. The origin of the peculiar behavior of the DB-SB model lies in the SB b.c.s defined in \eqref{ghsp}, which equips the ghost spins with an average magnetization close to $m_+$. On the other hand, because of the local equilibrium between the reservoir and the rightmost column that is produced by the detailed balance, ${\cal R}_+$  does not impose to  $\overline{m}_L$ its ``nominal''  magnetization $m_+$ (as with the ISF dynamics): the measured value of $\overline{m}_L$ is, indeed, typically larger than $m_+$ (cf. e.g. the plot referring to the DB-L4 dynamics). The outcome of the two competitive effects is shown in the bottom right panel of \figref{fig:mL}.

\begin{figure}
\centering
(a)\includegraphics[height=5.15cm]{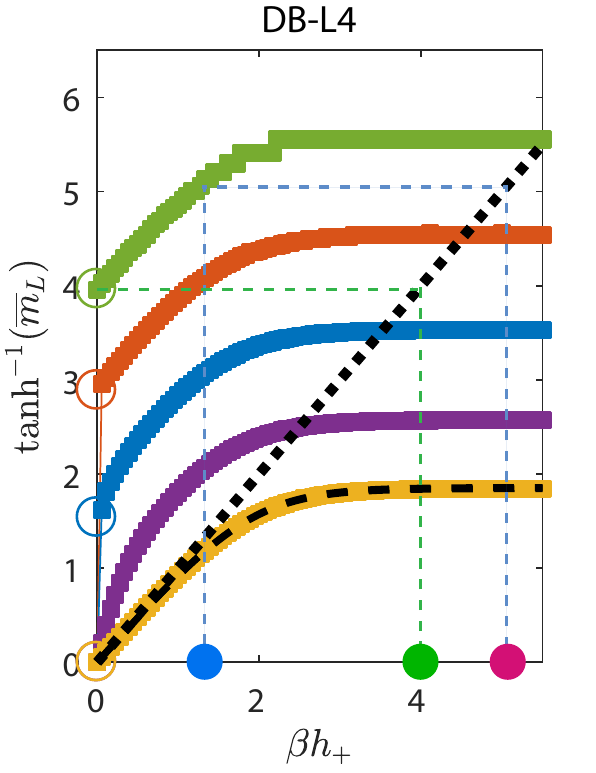}
(b)\includegraphics[height=4.85cm]{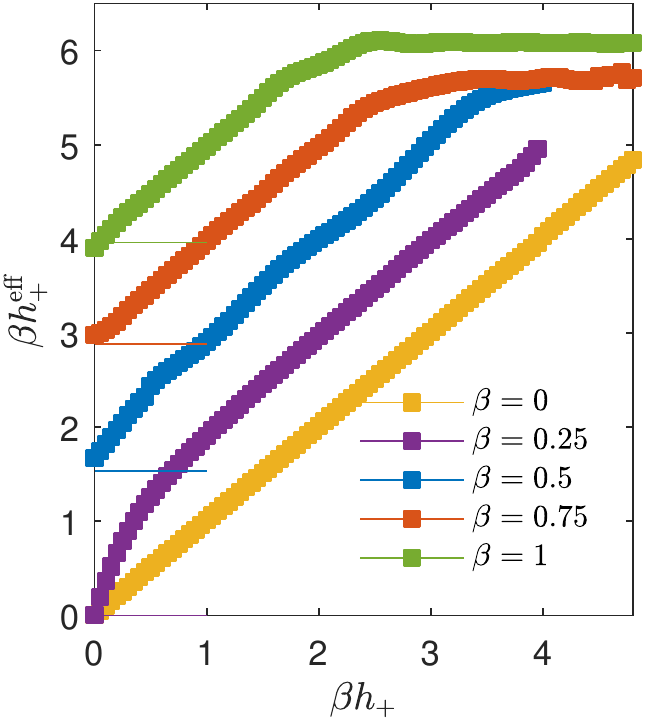}
\caption{{\bf (a)} Here we reproduce the bottom left panel  of \figref{fig:mL} for the mean magnetization at the boundary
to visualize the definition of $m^\textrm{eff}_+$ exemplified for the case of $\beta=1$ (green line). An arbitrary value of $m_+$ is singled out on the horizontal axis (blue bullet). The red bullet represents the ``effective magnetization'' $m^\textrm{eff}_+$. It yields, when used instead of $m_+$ during ISF-L4 dynamics, the same value $\overline{m}_L$ obtained with DB-L4 dynamics at $m_+$.
$\overline{m}_L^\textrm{ISF-L4}\approx m_+$ (dashed black line) holds for this rather large value of $\beta$.
$m^\textrm{eff}_+$ is larger than $m_\beta$ (green bullet).
{\bf (b}) Effective $m^\textrm{eff}_+$ versus $m_+$ for the same $\beta$'s.
Also included are values $m_\beta$ %$\mcrit^\textrm{ISF-L4}$
(short horizontal lines). We find that $m^\textrm{eff}_+>m_\beta$ for all $m_+$ and all $\beta$ investigated.
Because the calculation of $m^\textrm{eff}_+$ requires an inversion of $\overline{m}_L^\textrm{ISF-L4}$
with respect to $m_+$, the error bars are of the order of the visible undulations, and some data points at large $m_+$ have even been skipped.
ISF-L4 (and also DB-SB) dynamics equipped with $m^\textrm{eff}_+$ instead of $m_+$ exhibit just standard Fickian diffusion (results not shown).}
\label{fig:DB}
\end{figure}

Remarkably, only Fickian diffusion is observed with DB-L4 dynamics. To see that, let us first introduce the effective magnetization $m^\textrm{eff}_+$, defined as the reservoir magnetization which, with a ISF dynamics, gives rise to the same value of boundary magnetization $\overline{m}_L$ obtained with a DB dynamics (equipped with the same b.c.'s of the ISF dynamics) in presence of a reservoir magnetization $m_+$. Precisely, considering the L4 b.c.'s, one has that $m^\textrm{eff}_+$ solves the following relation:
\be
\overline{m}_L^\textrm{ISF-L4}(\beta,m_+^\textrm{eff}) = \overline{m}_L^\textrm{DB-L4}(\beta,m_+)
\label{meff}
\ee
Thus, \figref{fig:DB} reveals that, for any $\beta$, $m^\textrm{eff}_+$ is larger than $m_{\beta}$, which, as shown in \eqref{fig3.1}, is also close to $\mcrit^\textrm{ISF-L4}$. The result is, then, the onset of a stable phase, accompanied by a downhill current, cf. Section\ \ref{sec:results}.

Turning to the DB-SB model, the effective magnetization obtained from the solution of an equation similar to \eqref{meff} turns out being smaller, for certain values of $m_+$, than $\mcrit^\textrm{ISF-SB}$, in which case uphill currents can indeed be observed, see \figref{maps}.
Our results seem to indicate that purely Fickian diffusion is restored with the above DB-SB dynamics if one defines differently the SB boundary condition in \eqref{ghsp}, namely by replacing $m_+$ by $m^\textrm{eff}_+$ obtained from the solution \eqref{meff}, cf. the right panel of \figref{fig:DB} (the corresponding 2D map is not shown since it looks similar to the right upper panel of \figref{maps}).

Furthermore, the following argument allows us to interpret the behavior of $\overline{m}_L$ as a function of $m_+$ at $\beta=0$ for the ISF models. At $\beta=0$ all attempted bulk bond flips are accepted, $c_{i,j}^\textrm{bulk}=1$, and the dynamics is equivalent to that of a simple exclusion process \cite{simpleexclusion}. This implies that the magnetization profile $\overline{m}_x$ varies strictly linearly with $x$,  as it can be seen, for instance, by using duality \cite{CGGR2013}. The two coefficients characterizing the linear profile are obtained from the following two conditions: (i) For symmetry reasons $\overline{m}_1=-\overline{m}_L$, i.e., $\overline{m}_{(L+1)/2}=0$, and (ii) the magnetization at the boundary is determined with equal weight by bond flips from the adjacent layer with magnetization $\overline{m}_{L-1}$ and by the ISF updating scheme enforcing magnetization $m_+$. The resulting linear magnetization profile $\overline{m}_x$ and its values at the boundaries are
\bea
 \overline{m}_x = \frac{2x-1-L}{L+1}m_+, \qquad \overline{m}_L = -\overline{m}_1 = \frac{L-1}{L+1}m_+ \qquad (\beta=0)\, . \label{beta0theory}
\eea
As the case of $\beta=0$ provides a lower bound, this is the $\beta$-independent result for large system sizes. The observed departures reflect the finite system size and can actually be used to measure the system size.

In the following subsections we will present the results of a set of simulations at high  ($\beta=0.3$) and low ($\beta=1$) temperatures.  For both cases we will describe the asymptotic magnetization profile and the corresponding current.  These result are taken from a much larger bunch of simulations and are selected in order to offer an highlight of the complex dynamic presented by our models. As it could be expected, the
scenario below the critical  temperature  is much complex than above, where no anomalies occur.

\vskip .2cm

\subsection{High temperature regime}

\begin{figure}[tb]
\centering
\includegraphics[width=0.8\textwidth]{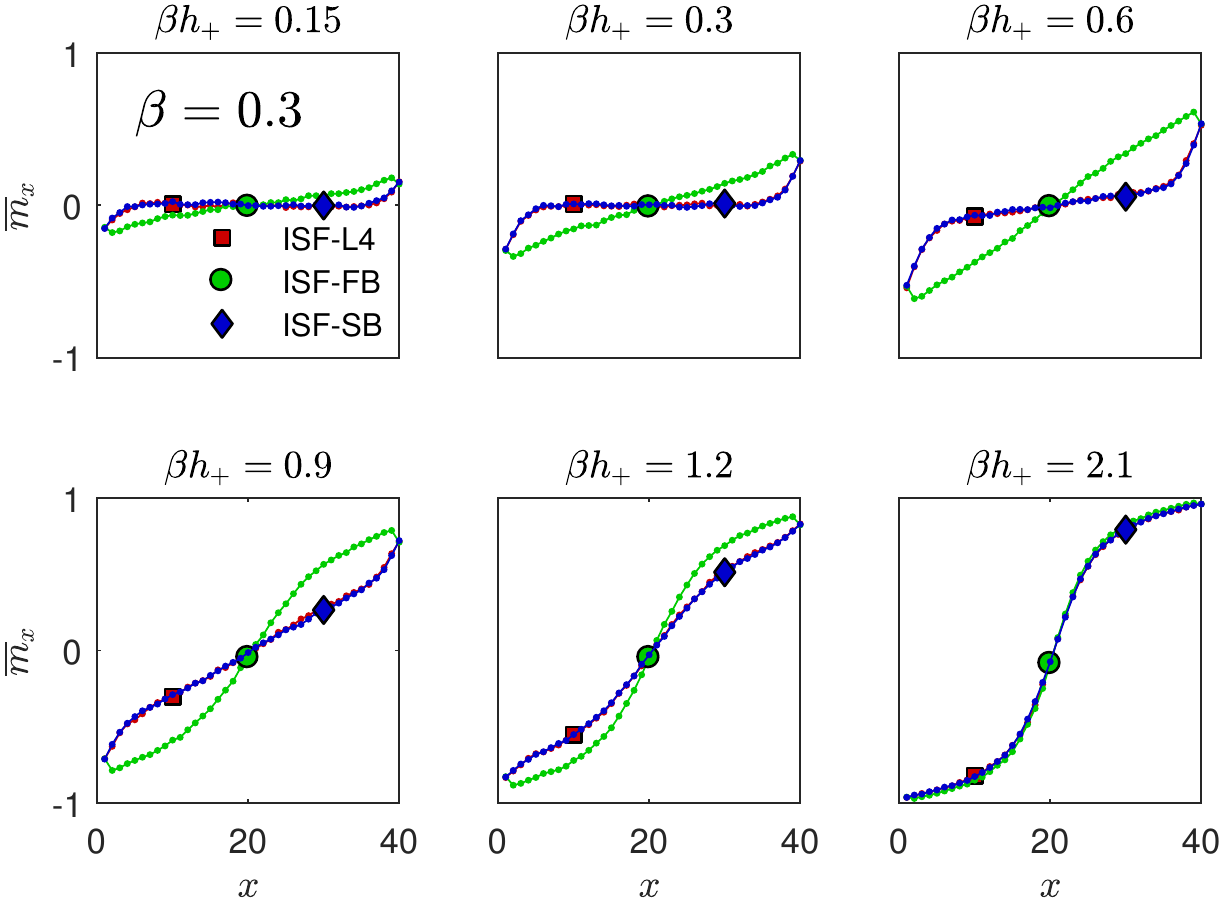}
\caption{\label{fig1.a1}Same as \figref{fig1.d1} for $\beta=0.3$, i.e. $\beta < \beta_c$.
A table of figures with the magnetization profiles $\overline{m}_x(\tau)$ versus $x$ for different values of $\betah_+$,
for the three nonequilibrium models.
}
\end{figure}

\begin{figure}
\centering
\includegraphics[width=0.7\textwidth]{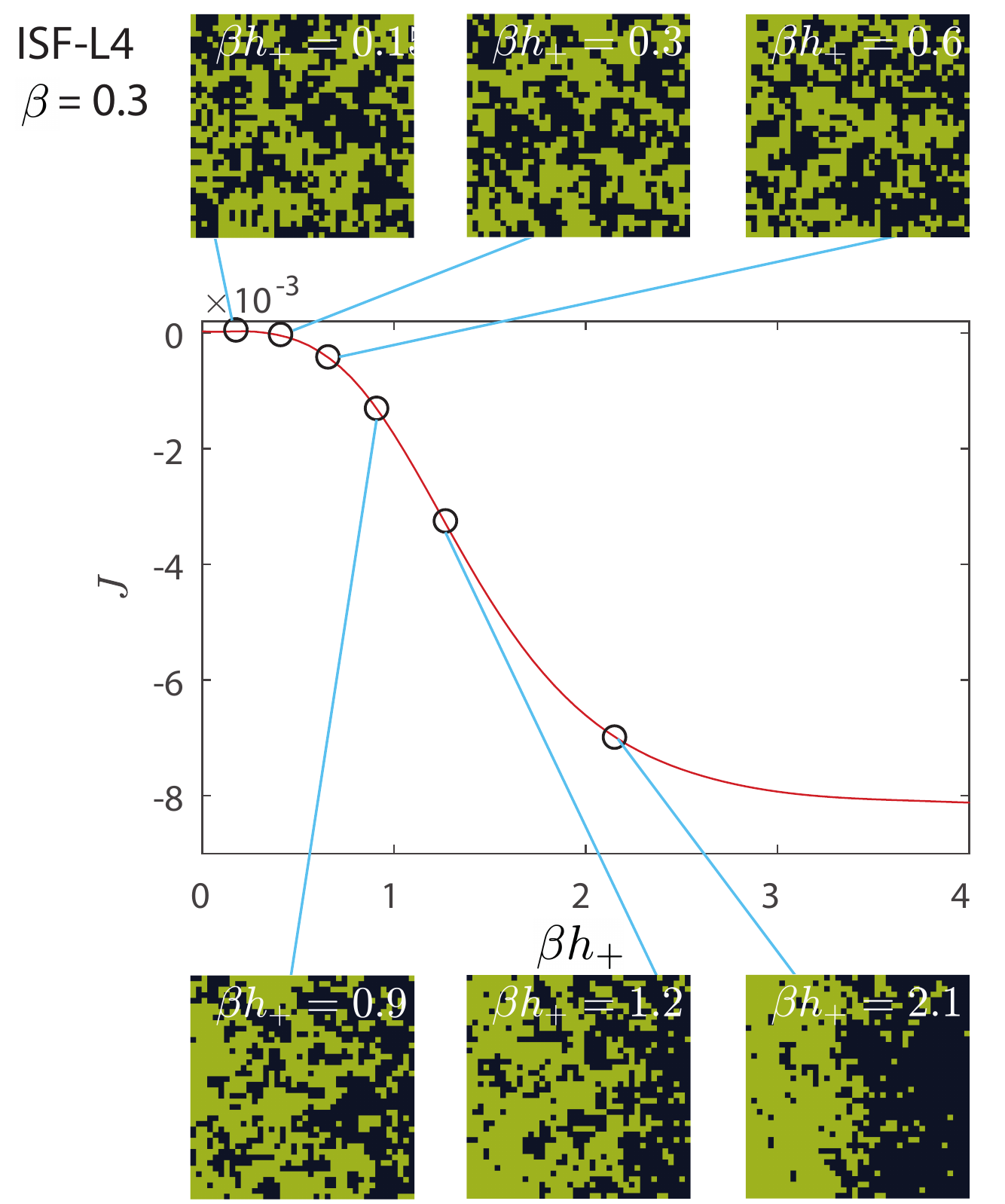}
\caption{\label{schematic-beta=03-ISF-L4}Nonequilibrium current $J$ versus dimensionless magnetic field $\betah_+$ for $\beta=0.3$ (ISF-L4), together with
selected spin configurations at time $t=\tau=10^{13}/N_b$. The corresponding magnetization profiles
are shown in \figref{fig1.a1} in concert with profiles for the other two nonequilibrium models (ISF-FB and ISF-SB). Color code: spin +1 (black), spin -1 (olive).}
\end{figure}

\begin{figure}
\centering
\includegraphics[width=0.6\textwidth]{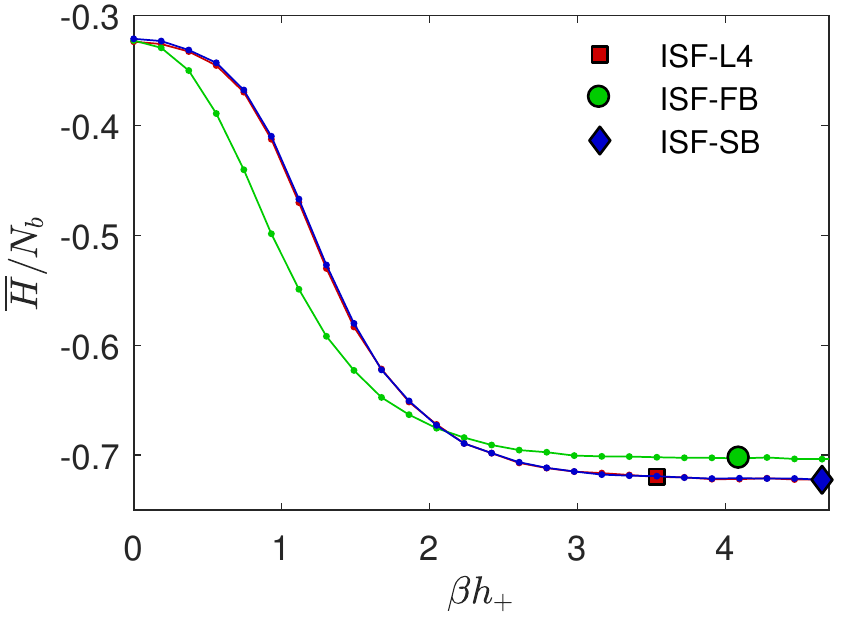}
\caption{\label{E-beta=03-ISF-L4}Mean total hamiltonian, as defined in \eqref{H-Ising}, per bond
vs $\betah_+$ for $\beta=0.3$, after $10^{13}$ MC steps for each $h_+$ value.
}
\end{figure}

We consider $\beta=0.3$ as a representative case of the high temperature regime.

In \figref{fig1.a1}  we display the stationary magnetization profiles  $\overline{m}_{x}$ for different values of $\betah_+$. In each panel the three models ISF-L4, ISF-FB and ISF-SB are considered.
As it is evident from the figure, the profiles of the three different b.c.s are qualitatively comparable in that they all are  non decreasing, at least for not too small $\betah_+$'s, say for $\betah_+\ge 0.6$.  However, the b.c.s play a role: while the
L4 and SB are essentially indistinguishable in all the range of $\betah_+$'s, the FB coincides with the others only at high $\betah_+$, including $m_+=1$, but decreasing $\betah_+$ the curvature of FB profile  appears clearly different from the others. For $\betah_+<0.6$ the L4 and SB curves develop, eventually through the appearance of slightly pronounced local maxima and minima, an almost flat region in the bulk.  A similar
behavior is developed also for the FB case, but for smaller $\betah_+$.
Such profiles sustains a negative current, as we can see in \figref{schematic-beta=03-ISF-L4} where $J$ is plotted versus $\betah_+$. Moreover, $J$ looks as a smooth curve that decreases monotonically as $\betah_+$ increases from $0$.
As $\betah_+\to 0$ the systems get close to equilibrium, being $m_+$ and $m_-=-m_+$ close since $m_+ \to 0$. Thus $\overline{m}_{x}$ approaches the flat equilibrium profile, as observed above, and the flux approaches 0.

The analysis of the corresponding spin configurations shows a weak phase separation (without sharp interface, see the slope in the center of the curve for $\betah_+=2.1$) that becomes less and less visible approaching  $\betah_+=0$. This indicates that, as the system approaches equilibrium, the positive and negative phases become more and more random and intertwined: the configuration loses the instanton profile structure.
The confirmation of the fact that as $\betah_+$ decreases, the spin configuration becomes more disordered is shown in \figref{E-beta=03-ISF-L4}, where the mean total Hamiltonian per bond $\overline{H}/N_b$ is represented for the ISF-L4, ISF-FB and ISF-SB models. Since the Hamiltonian can be written as $H(\sigma|\sigma^0)=2N_{-}-N_b$, where $N_{-}$ is the number of broken bonds (a broken bond has antiparallel spins),
this quantity, which is in between $-1$ and $1$, gives a measure of the amount of broken bonds on the total lattice bonds.  Its value  is  -1 when no bond is broken and 1 when all the bonds are broken, while for a sharp vertical interface (with essentially $L$ broken bonds), this ratio is $(1-2L)/(1+2L)$, whose value is $\approx -0.975$  for $L=40$.  Also in this respect the L4 and SB models are the same,  while FB follows a different path, though qualitatively similar.

\subsection{Low temperature regime}

As a case study for the low temperature regime, we consider $\beta=1$,
and write $h_+$ instead of $\beta h_+$ in this section, to simplify notation.

The analysis of the time averaged magnetization profiles shown in \figref{fig1.d1}, offers a rich scenario as $\skipbetah_+$ decreases: starting from the instanton profile, which  is common to all three ISF models  in  the stable parameter region (see panel for $\skipbetah_+=5.35$), we find a metastable phase where the instanton is replaced by the bump (see panel for $\skipbetah_+=3.7$). Then continuing to decrease $\skipbetah_+$, the system enters into the weakly unstable phase with double-bump profiles (see panel for $\skipbetah_+=2.25$). In this case two (almost) flat regions with values close to $m_+$ and $m_-$ develop  in the vicinity of the boundary with the opposite magnetization. Then moving towards the boundary, the profile of $\overline{m}_x$ jumps sharply to the value imposed by the reservoir. This behavior corresponds to a configuration with an interface in the middle, but with the plus and minus phases flipped, i.e.  close respectively  to ${\cal R}^-$ and ${\cal R}^+$. Because of that, two boundary layers appear. A configuration with this behavior is clearly visible, in \figref{schematic-beta=1-ISF-L4} for
$\skipbetah_+=2.24$. Decreasing $\skipbetah_+$ further, sharper maxima and minima appear, see e.g. $\skipbetah_+=1.05$.

Observe that for a given $\skipbetah_+$, while the
behaviors of ISF-L4 and ISF-SB models are similar (maybe after the application of the left-right and spin flip symmetries, see e.g.  $\skipbetah_+=1.35$), the free b.c.s of the ISF-FB model generate quite different nonequilibrium stationary states. However, as in the high temperature case, the three curves seem to reconcile as $\skipbetah_+\to 0$ (that is approaching the equilibrium)  when in the bulk the profile tend to flatten, although with two peaks close to the reservoir that resemble those found in Rieder et al.\ \cite{Rieder1967} or the widely observed boundary resistance effects.

\begin{figure*}
\centering
\includegraphics[width=0.55\textwidth]{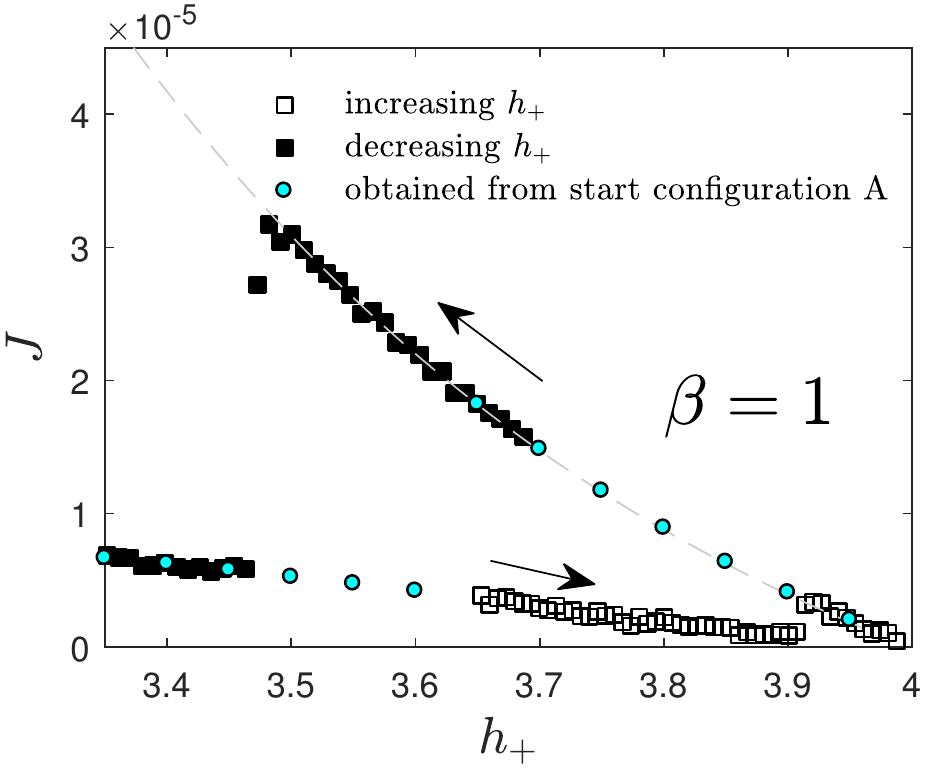}
\caption{\label{fig-hysteresis-beta=1-run1}Zoom into results for $J$ at $\beta=1$. A more detailed investigation of the
hysteretic region at $\skipbetah_+$ slightly below the critical $\skipbetah_{\beta=1}\approx 3.96158$, where results depend on the initial configuration.
Cyan circles mark results obtained after $\tau=10^{13}/N_b$ starting from the default initial configuration A.
The open squares are obtained by slowly increasing $\skipbetah_+$ with a positive rate equal to $6.690\times10^{-14}$ per MC step, starting from the cyan configuration at $\skipbetah_+=3.6$,
the filled squares are obtained by slowly decreasing $\skipbetah_+$ with a negative rate equal to $-9.315\times10^{-14}$ per MC step, starting from the 'cyan' configuration at $\skipbetah_+=3.7$.}
\end{figure*}

\begin{figure}
\centering
\includegraphics[width=0.5\textwidth]{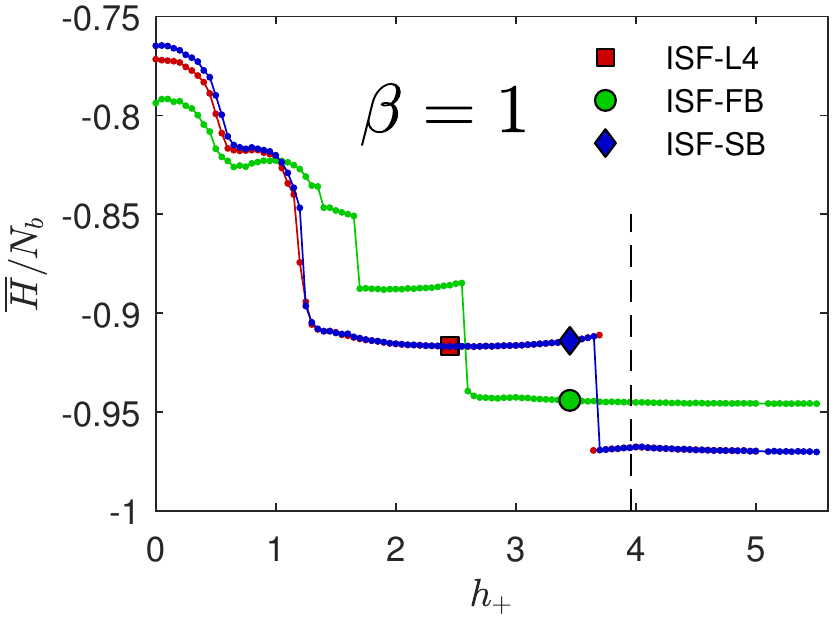}
\caption{\label{E-beta=1-ISF-L4}Mean total hamiltonian as defined in \eqref{H-Ising} divided by number of bonds
vs $\skipbetah_+$ for $\beta=1$.
The dashed line marks the Onsager $\skipbetah_{\beta=1}\approx 3.9616$ according to \eqref{mbeta}.
In the limit of $\skipbetah_+\rightarrow\infty$ ($m_+=1$) the mean energy $\overline{H}/N_b$ approaches
the energy of the default initial configuration A with a single straight line interface $\overline{H}/N_b=(1-2L)/(1+2L)\approx -0.975$, while there are clearly about three line interfaces for $\skipbetah_+\in[1.5,3.6]$, the ones visible in \figref{schematic-beta=1-ISF-L4}.
$40\times 40$ grid.
$\tau=10^{13}/N_b$ for each $\skipbetah_+$.
}
\end{figure}

\figureref{schematic-beta=1-ISF-L4} shows the current for ISF-L4 model: in contrast with the simplest high temperature case, in this low temperature region the current curve, reflecting the complexity of the magnetization profiles, is  not smooth and  not monotonic.  It  crosses three times the zero line and shows the presence of an hysteretic region (close to $\skipbetah_+=4$) where the observed stationary state depends on the initial configuration chosen in the MC simulation. A blow up of this  phenomenon is shown  in \figref{fig-hysteresis-beta=1-run1}. Cyan circles mark results obtained after $\tau=10^{13}/N_b$ MC steps starting from the default initial configuration A.
The open squares are obtained by slowly \emph{increasing} $\skipbetah_+$, starting from the cyan configuration at $\skipbetah_+=3.6$,
while the filled squares are obtained by slowly \emph{decreasing} $\skipbetah_+$, starting from the 'cyan' configuration at $\skipbetah_+=3.7$. At these rates the hysteretic region is seen to extend over the range $\skipbetah_+\in[3.46,3.91]$.

The spin configurations corresponding to the magnetization profiles of ISF-L4  (represented by red square lines in \figref{fig1.d1}) are shown in the surrounding panels of \figref{schematic-beta=1-ISF-L4}. Note that the study of the mean total hamiltonian per bond $\overline{H}/N_b$ shows, see \figref{E-beta=1-ISF-L4},  wide intervals in which the curves are almost constant, meaning that in those regions the interfaces separating the plus and minus phase are maintained. It is worth noting the difference with the high temperature case, see \figref{E-beta=03-ISF-L4}, in which $\overline{H}/N_b$ varies smoothly without flat regions.

\begin{figure*}[tb]
\centering
(a)\includegraphics[width=0.46\textwidth]{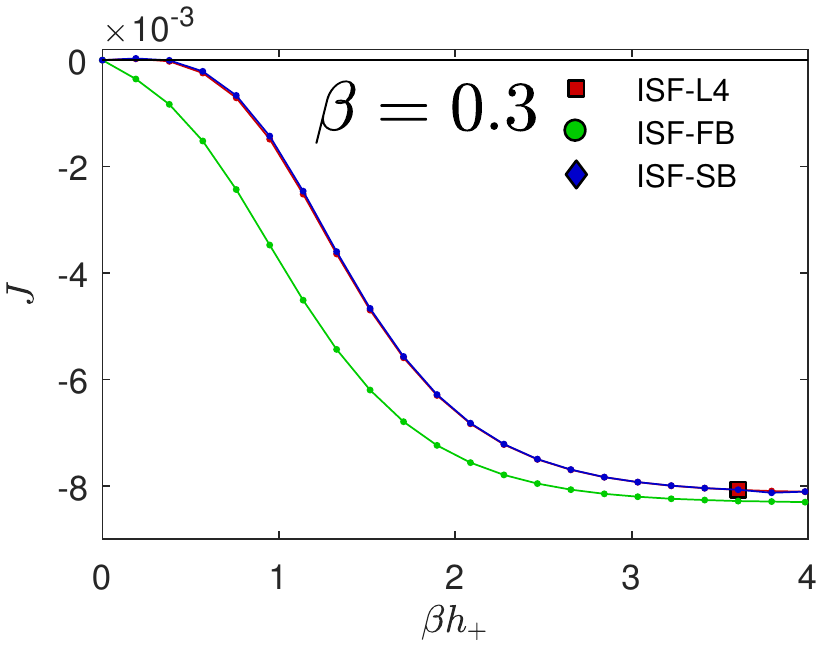}
(b)\includegraphics[width=0.46\textwidth]{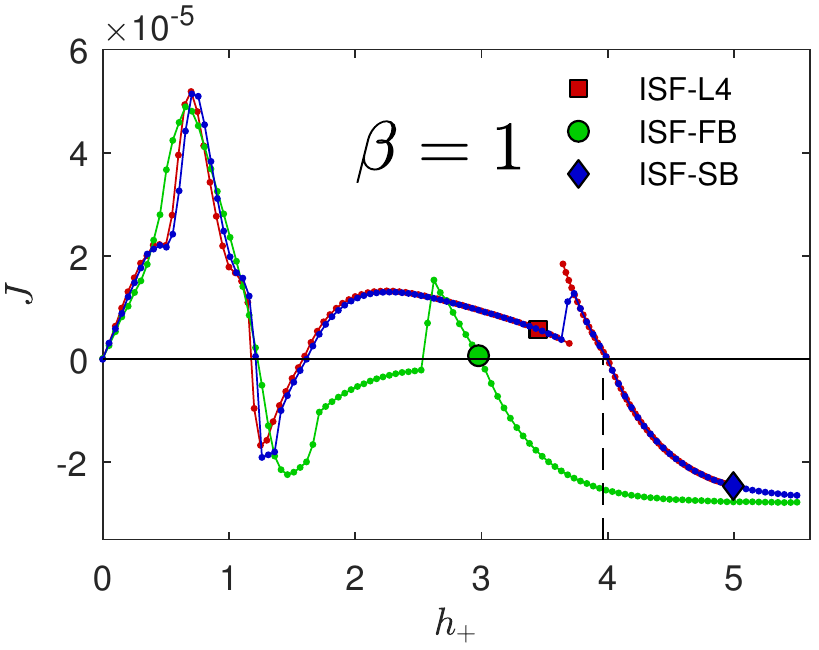}
\caption{\label{fig1.e}Current $J$ vs $\betah_+$ reporting the different curves corresponding to the considered values of (a) $\beta=0.3$ and (b) $\beta=1.0$.
The dashed line in (b) $\beta=1$ marks the Onsager $h_{\beta=1}\approx 3.9616$ according to \eqref{mbeta}.
Each point marks an independent simulation with $\tau=10^{13}/N_b$. $40\times 40$ grid. Initial configuration A. The red curves (ISF-L4) had
already been shown together with configurations in \figsref{schematic-beta=1-ISF-L4}{schematic-beta=03-ISF-L4}.
Note the different vertical scales in (a) and (b).
}
\end{figure*}

\begin{figure}
\centering
\includegraphics[width=0.5\textwidth]{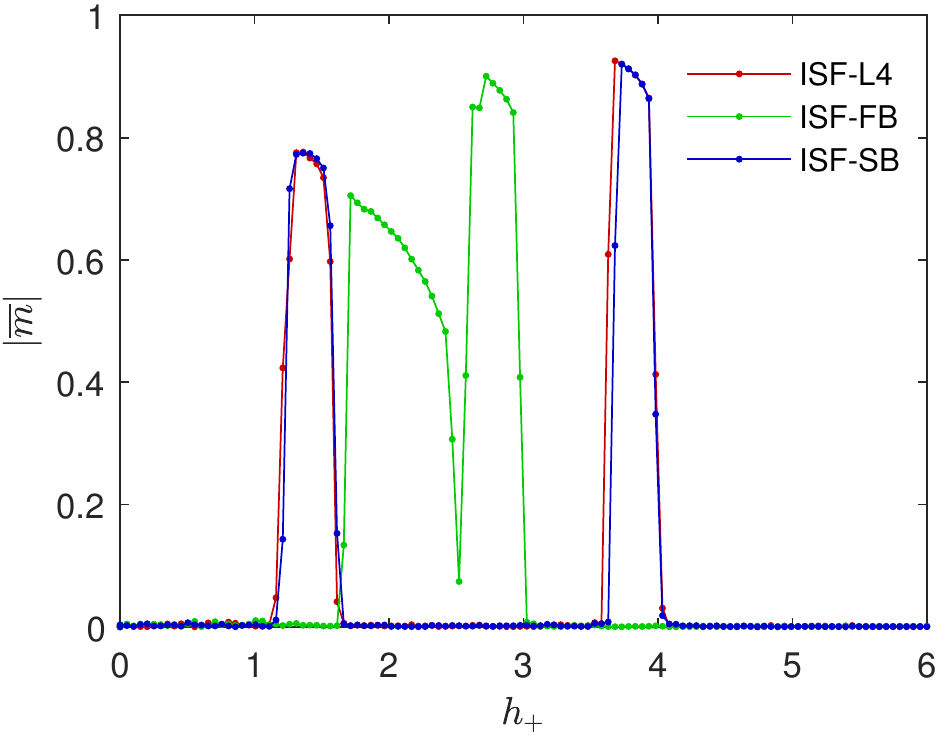}
\caption{\label{fig-m}Absolute mean total magnetization $|\overline{m}|$ vs $h_+$ for $\beta=1$ (at $\tau=10^{13}/N_b$), obtained from the magnetizations
$m(\sigma)$ defined in \eqref{magn}, or alternatively, the averaged magnetization profiles $\overline{m}_x$ defined after \eqref{defmx}. For $\beta=0.3$ the mean magnetization vanishes for all $h_+$.}
\end{figure}

\figureref{fig1.e} represents the current $J$ as a function of $\betah_+$ for the two cases $\beta=0.3$ (left panel) and $\beta=1$ (right panel). Together with the current curves already shown in \figsref{schematic-beta=03-ISF-L4}{schematic-beta=1-ISF-L4} for ISF-L4 model, there are the ones for  ISF-FB and ISF-SB models. From the comparison it is confirmed that also for this quantity the model ISF-SB behaves like the model ISF-L4, while ISF-FB model differs from the first two, while maintaining qualitatively similar trends within the same temperature regime. Again, the difference between high and low temperature cases is evident.
\figureref{fig-m} shows that at low temperature ($\beta=1$) there are intervals of $\betah_+$ where the absolute value of the average magnetization $|\overline{m}|$ is non-zero, while for high temperature the mean magnetization vanishes for all $\betah_+$.

\section{Conclusions}
\label{sec:concl}

We have investigated the behavior of the current $J$ and the boundary magnetization $\overline{m}_L$ for some 2D Ising models, coupled to external magnetization reservoirs and equipped with different b.c.s, called respectively L4, FB and SB.

In particular, we examined the sign of the stationary currents in the parameter space, spanned by the variables $\beta$ and $m_+$. Our MC simulations indicate that stationary uphill currents do appear for certain values of the parameters of the model, as a result of the ISF spin-updating mechanism, which breaks the condition of detailed balance. We also highlighted the relation between $\mcrit$ and $\meq$, the first being an observable characterizing the nonequilibrium dynamics of the 2D Ising model in contact with external reservoirs, while the latter refers to an equilibrium Ising model with conservative dynamics. Our simulations show that, if $L$ is fixed, the two quantities both tend to Onsager's magnetization $m_{\beta}$ when $\beta$ is large.

Moreover, we studied in detail the behavior of the equation of state, namely the relation $J$ vs. $m_{+}$, for $\beta=0.3$ and $\beta=1$, as representative, respectively, of the high temperature and low temperature regimes. Lowering the parameter $m_+$ leads to novel stationary states, in which the current changes sign multiple times and more interfaces appear in the microscopic spin configurations.
One open question, not addressed in this manuscript, concerns the presence of uphill diffusion in the infinite volume limit. More theoretical work is needed to clarify whether uphill currents persist in the above limit.

The analysis of 2D Ising models on a finite lattice, coupled to external reservoirs that break the condition of detailed balance, may be relevant in a variety of applications, e.g. in the investigation of mesoscopic systems, in which the notion of ``local equilibrium'' is not guaranteed. We thus expect uphill currents to play a major role for these systems, that do not follow the basic tenets of irreversible thermodynamics.

\paragraph*{Acknowledgments.}
The authors wish to thank Anna De Masi (University of L'Aquila) and Errico Presutti (Gran Sasso Science Institute) for inspiring this work and for the many useful comments and discussions.
C.~Giardin\`{a} is acknowledged for useful discussion.
MC acknowledges financial support from FFABR 2017.
CG and CV acknowledge financial supports
from Fondo di Ateneo per la Ricerca 2016 and 2017 (UniMoRe).

\bibliographystyle{plain}      % American Physical Society (APS) style, author-year citations
\bibliography{references}
\nocite{*}

\end{document}